\documentclass[12pt]{article}
\pdfoutput=1
\usepackage{jheppub}
\usepackage[utf8]{inputenc}
\usepackage{verbatim}
\usepackage{amsmath}
\usepackage{amssymb}
\usepackage{amsthm}
\usepackage{slashed}
\usepackage{amsfonts}
\usepackage{subfigure}
\usepackage{float}
\usepackage{physics}
\usepackage[capitalize]{cleveref}

\makeatletter
\def\blfootnote{\gdef\@thefnmark{}\@footnotetext}
\makeatother

\preprint{CALT-TH-2020-024, DMUS-MP-20/11, YITP-20-121}

\begin{document}

\title{Quantum simulation of gauge theory via orbifold lattice}
\author[a]{Alexander J. Buser,}
\author[a]{Hrant Gharibyan,}
\author[b,c]{Masanori Hanada,}
\author[c]{Masazumi Honda,}
\author[a]{Junyu Liu}

\affiliation[a]{Walter Burke Institute for Theoretical Physics and Institute for Quantum Information and Matter, California Institute of Technology, Pasadena, CA 91125, USA}
\affiliation[b]{Department of Mathematics, University of Surrey, Guildford, Surrey, GU2 7XH, UK}
\affiliation[c]{Yukawa Institute for Theoretical Physics, \\ Kyoto University, Kitashirakawa Oiwakecho, Sakyo-ku, Kyoto 606-8502, Japan}
\let\thefootnote\relax\footnotetext{\textbf{Authors are alphabetically ordered.}}

\abstract{
We propose a new framework for simulating $\text{U}(k)$ Yang-Mills theory on a universal quantum computer. 
This construction uses the orbifold lattice formulation proposed by Kaplan, Katz, and Unsal, who originally applied it to supersymmetric gauge theories. 
Our proposed approach yields a novel perspective on quantum simulation of quantum field theories, carrying certain advantages over the usual Kogut-Susskind formulation. 
We discuss the application of our constructions to computing static properties and real-time dynamics of Yang-Mills theories, from glueball measurements to AdS/CFT, making use of a variety of quantum information techniques including qubitization, quantum signal processing, Jordan-Lee-Preskill bounds, and shadow tomography. 
The generalizations to certain supersymmetric Yang-Mills theories appear to be straightforward, providing a path towards the quantum simulation of quantum gravity via holographic duality. 
}

\maketitle

\section{Introduction}
\hspace{0.51cm}
Quantum simulation is among the most important applications of quantum hardware, both for near-term and fault-tolerant quantum computation, addressing the capability of quantum devices to probe, calculate, and simulate real problems appearing in the physical world. Theoretically, constructing efficient quantum simulation algorithms enhances support for the claim of the quantum Church-Turing Thesis, which states that one can simulate all physical processes using quantum circuits with reasonable costs in time. 
Practically, quantum simulation of fundamental physical processes may shed light on complex phenomena appearing in quantum gravity, quantum cosmology, sub-atomic particle physics, condensed-matter physics, cold-atomic physics, and statistical physics where classical computers encounter insurmountable challenges. 
We are in an era of quickly developing quantum technology, where near-term quantum computers may perform tasks surpassing the current capabilities of classical computation. 
In the long-term, it is reasonable to expect that quantum devices will perform universal fault-tolerant quantum computation, allowing us to run quantum algorithms reliably
(see some general introduction about this topic in \cite{preskill2018quantum,Preskill:2018fag,cyber}).

Out of the many applications of quantum simulation, implementing quantum field theory is a particularly promising area for realizing a useful quantum advantage. Already, detailed studies have been carried out demonstrating the efficiency of simulating the dynamics of scalar field theories \cite{Preskill:2018fag,Jordan:2011ne,Jordan:2011ci,Klco:2018zqz}. 
Many outstanding problems in quantum field theory addressable by quantum computation concern the properties of lattice gauge theories.
Gauge theories are not only the foundation of particle physics and condensed-matter physics, but also play a critical role in quantum error correcting codes and topological quantum computation. 
For instance, the celebrated toric code developed by Kitaev is naturally understood in terms of a $\mathbb{Z}_2$ lattice gauge theory. 
Much attention has been given recently to the task of simulating lattice gauge theory on a quantum computer. Most proposals use the traditional Hamiltonian formulation of lattice gauge theory developed by Kogut and Susskind\footnote{Another promising approach is to start directly from a finite-dimensional model which possesses a continuum limit in the universality class of the desired quantum field theory \cite{PhysRevD.100.054505,buser2020quantum,PhysRevD.60.094502,CHANDRASEKHARAN1997455}.} (KS) \cite{Kogut:1974ag}. 
Yet another Hamiltonian formulation is readily obtained from the lattice Yang-Mills action based on the {\it orbifold construction} introduced by Kaplan, Katz, and Unsal (KKU) in 2002 \cite{Kaplan:2002wv}. 
In this paper, we will study how the orbifold construction can be used for digital quantum simulation and analyze its advantages and disadvantages compared to an approach based on the KS formulation.

The original motivation for the orbifold construction was to realize supersymmetry on a lattice; 
by performing an orbifold projection on a supersymmetric Yang-Mills matrix model, one obtains a lattice theory preserving a few supersymmetries, which is (for $(1+1)$ and $(1+2)$ dimensions) sufficient for the complete restoration of all supercharges in the continuum limit. 
This idea was subsequently applied to Euclidean theories \cite{Cohen:2003xe,Cohen:2003qw,Kaplan:2005ta} 
and inspired alternative approaches to realizing supersymmetry on a Euclidean lattice, e.g. Refs.~\cite{Sugino:2003yb,Sugino:2004qd,Sugino:2004uv,Catterall:2003wd,Catterall:2004np}. 
In these formulations, no parameter fine tuning is needed to achieve the supersymmetric continuum limit to all orders in perturbation theory.
Numerical results \cite{Hanada:2009hq,Hanada:2010qg,Catterall:2011aa,Giguere:2015cga} further support the expectation that fine tuning is not necessary even at the nonperturbative level. 

Therefore, the orbifold construction has a clear advantage when applied to supersymmetric theories. 
As far as Euclidean theories are concerned, there may be no clear motivation for applying the orbifold construction to non-supersymmetric theories, as traditional lattice regularizations such as Wilson's plaquette action \cite{Wilson:1974sk} are well understood and already sufficiently useful. 
However, in this paper, we will show that the situation is different for real-time quantum simulation. 
In fact, the orbifold construction could provide important, non-substitutable inputs when simulating lattice gauge theories in a quantum computer. 
Although we consider $\text{U}(k)$ gauge theory as a specific example, 
we expect that our method can be generalized to other gauge groups useful for, for instance, studies of the Standard Model in particle physics.\footnote{
The Standard Model introduces other challenges; non-perturbative regularizations of chiral fermions are notoriously difficult, for instance. For the vector-like theories (i.e., left- and right-handed sector appear together such as in QCD), known methods on a lattice exist, and might be applicable to quantum simulation, for instance,  
domain wall fermions~\cite{KAPLAN1992342} and the 
overlap fermions~\cite{NEUBERGER1998141}.
}

The original motivation of the orbifold lattice construction was the study of supersymmetric Yang-Mills theory, and more specifically, quantum gravity via holographic duality. 
It appears that the construction we discuss in this paper can be generalized to supersymmetric theory. 
Therefore, together with a recent paper \cite{matrixModel}, which studied the approach based on the matrix models, this paper may serve as the first step towards quantum simulation of deep problems in quantum gravity via holography, such as the black hole information puzzle and emergent spacetime. 

This paper is organized as follows. 
In \Cref{sec:construction}, we review the orbifold construction and discuss its Hamiltonian version. 
In \Cref{sec:realization_on_QC}, we show how to realize this model as quantum simulation, particularly for the task of preparing ground states and measuring observables.  
In \Cref{sec:comparison}, we compare the orbifold and Kogut-Susskind approaches as platforms for quantum simulation. 
In \Cref{sec:conclusion}, we present some topics for future work.
In Appendix~\ref{sec:orbifold-projection}, we explicitly show how the lattice theory we consider is obtained from an orbifold projection on a matrix model. 
In Appendix~\ref{sec:Kogut-Susskind}, we review the Kogut-Susskind formulation of lattice gauge theory. Appendix~\ref{sec:coordinate_basis} contains some notes on an alternative digitization scheme, which may be more practical than that considered in the main text in some cases.

\section{Orbifold construction of lattice gauge theory}\label{sec:construction}
\hspace{0.51cm}
In this section, we introduce the orbifold construction of pure Yang-Mills theory on a lattice. 
The field content and Lagrangian are explained in Sec.~\ref{sec:orbifold-lattice-explicit-construction}. 
The orbifold construction uses non-compact variables rather than compact variables (unitary link variables); 
in Sec.~\ref{sec:KKU-vs-Wilson-Lagrangian}, the relation to the formulation with unitary link variables is made clear. 
The Hamiltonian formulation, which is used for the implementation on a quantum computer 
in later sections, is introduced in Sec.~\ref{sec:operator-formalism}. 
The symmetry of the orbifold lattice at the discretized level is examined in Sec.~\ref{sec:symmetry-orbifold-lattice}. 

The adjective `orbifold' comes from the original construction \cite{Kaplan:2002wv}, which obtained the lattice action from a matrix model via the orbifold projection. We review the details of this construction in Appendix~\ref{sec:orbifold-projection}.\footnote{
The orbifold construction played a key role to find a supersymmetric lattice. 
For the purpose of this paper, the use of the orbifold projection is not crucial;  
just by accepting the action given in Sec.~\ref{sec:orbifold-lattice-explicit-construction} as a starting point, it is possible to understand how the orbifold lattice serves as a lattice regularization of Yang-Mills theory. 
}

For concreteness we consider the $(3+1)$-dimensional theory. Essentially the same construction works for $(2+1)$- and $(1+1)$-dimensional theories as well, as we will briefly see in the end of Sec.~\ref{sec:orbifold-lattice-explicit-construction}. 
\subsection{Orbifold lattice}
\label{sec:orbifold-lattice-explicit-construction}
\hspace{0.51cm}
The `orbifold lattice' version of U($k$) Yang-Mills is given as follows. 
We introduce a label of lattice points $\vec{n}=(n_x,n_y,n_z)$, where $n_x,n_y,n_z=1,2,\cdots,L$.
Then the `orbifold lattice' is the gauged matrix quantum mechanics 
with the gauge group $\prod_{\vec{n}}\text{U}(k)_{\vec{n}}$
which contains $k\times k$ complex matrices $x_{\vec{n}}$, $y_{\vec{n}}$ and $z_{\vec{n}}$
living on the links connecting $\vec{n}$ and $\vec{n}+\hat{x}$, $\vec{n}+\hat{y}$ and $\vec{n}+\hat{z}$, respectively.
As we will see shortly, the unitary link variables come out of these complex matrices.
We use a bar to denote Hermitian conjugate, i.e., $\bar{x}=x^\dagger$.
Then $\bar{x}_{\vec{n}}$, $\bar{y}_{\vec{n}}$ and $\bar{z}_{\vec{n}}$ are regarded as the link field with the opposite direction, 
 i.e., from $\vec{n}+\hat{x}$, $\vec{n}+\hat{y}$ and $\vec{n}+\hat{z}$ to $\vec{n}$, respectively\footnote{
In group theory language,
$x_{\vec{n}}$, $y_{\vec{n}}$ and $z_{\vec{n}}$ are bi-fundamental representations
of $\text{U}(k)_{\vec{n}}\times \text{U}(k)_{\vec{n}+\hat{x}}$, $\text{U}(k)_{\vec{n}}\times \text{U}(k)_{\vec{n}+\hat{y}}$ and $\text{U}(k)_{\vec{n}}\times \text{U}(k)_{\vec{n}+\hat{z}}$, respectively.
Their bars belong to the anti-bi-fundamental representations.
$A_{\vec{n}}$ is the gauge field for $\text{U}(k)_{\vec{n}}$. 
See \eqref{gauge-transf-scalar} and \eqref{gauge-transf-A} for the explicit definition of the gauge transformation.
}. 
We also add the gauge field $A_{\vec{n}}$ living on each site $\vec{n}$, which will be identified with the temporal component of the gauge field in $(1+3)$-d theory.
The Lagrangian is given by 
\begin{eqnarray}
L_{\rm lattice}
&=&
\sum_{\vec{n}}
{\rm Tr}\Biggl(
|D_tx_{\vec{n}}|^2 +|D_ty_{\vec{n}}|^2 +|D_tz_{\vec{n}}|^2 \nonumber\\
&& \quad
-\frac{g_{\rm 1d}^2}{2}\left|
x_{\vec{n}} \bar{x}_{\vec{n}} -\bar{x}_{\vec{n}-\hat{x}}x_{\vec{n}-\hat{x}}
+y_{\vec{n}}\bar{y}_{\vec{n}}
-\bar{y}_{\vec{n}-\hat{y}}y_{\vec{n}-\hat{y}}
+z_{\vec{n}}\bar{z}_{\vec{n}}
-\bar{z}_{\vec{n}-\hat{z}}z_{\vec{n}-\hat{z}} \right|^2 \nonumber\\
&& \quad
-2g_{\rm 1d}^2
\left( \left|
x_{\vec{n}}y_{\vec{n}+\hat{x}}
-y_{\vec{n}}x_{\vec{n}+\hat{y}}\right|^2
+\left| y_{\vec{n}} z_{\vec{n}+\hat{y}}
-z_{\vec{n}}y_{\vec{n}+\hat{z}} \right|^2
+\left| z_{\vec{n}} x_{\vec{n}+\hat{z}}
-x_{\vec{n}}z_{\vec{n}+\hat{x}} \right|^2 \right) \Biggl).  \nonumber\\
\label{eq:lattice-action}
\end{eqnarray}
Here we have used the notation $|M|^2=MM^\dagger$ for any matrix $M$. 
The trace ${\rm Tr}$ is over a $k\times k$ matrix.
The covariant derivative $D_t$ is defined by
\begin{eqnarray}
D_tx_{\vec{n}}
&=&
\partial_tx_{\vec{n}}
-
iA_{\vec{n}}x_{\vec{n}}
+
ix_{\vec{n}}A_{\vec{n}+\hat{x}}, 
\nonumber\\
D_ty_{\vec{n}}
&=&
\partial_ty_{\vec{n}}
-
iA_{\vec{n}}y_{\vec{n}}
+
iy_{\vec{n}}A_{\vec{n}+\hat{y}}, 
\nonumber\\
D_tz_{\vec{n}}
&=&
\partial_tz_{\vec{n}}
-
iA_{\vec{n}}z_{\vec{n}}
+
iz_{\vec{n}}A_{\vec{n}+\hat{z}}.
\end{eqnarray}
 A local U($k$) gauge transformation at site $\vec{n}$ parametrized by $\Omega_{\vec{n}}$ is given by 
\begin{eqnarray}
& &
x_{\vec{n}}
\to
\Omega_{\vec{n}}x_{\vec{n}}\Omega^{-1}_{\vec{n}+\hat{x}}, 
\qquad
y_{\vec{n}}
\to
\Omega_{\vec{n}}y_{\vec{n}}\Omega^{-1}_{\vec{n}+\hat{y}}, 
\qquad
z_{\vec{n}}
\to
\Omega_{\vec{n}}z_{\vec{n}}\Omega^{-1}_{\vec{n}+\hat{z}}, 
\label{gauge-transf-scalar}
\end{eqnarray}
and
\begin{eqnarray}
A_{\vec{n}}
\to
\Omega_{\vec{n}}A_{\vec{n}}\Omega^{-1}_{\vec{n}}
+
ig_{\rm 1d}^{-1}\Omega_{\vec{n}}\partial_t\Omega^{-1}_{\vec{n}}. 
\label{gauge-transf-A}
\end{eqnarray}
If we take the $A=0$ gauge, 
then the Gauss-law constraint coming from $\frac{\partial L}{\partial A}=0$ is given by 
\begin{eqnarray}
\sum_{\mu=1}^3
\left(
-
x_{\mu,\vec{n}}\dot{\bar{x}}_{\mu,\vec{n}}
+
\dot{x}_{\mu,\vec{n}}\bar{x}_{\mu,\vec{n}}
-
\bar{x}_{\mu,\vec{n}-\hat{\mu}}\dot{x}_{\mu,\vec{n}-\hat{\mu}}
+
\dot{\bar{x}}_{\mu,\vec{n}-\hat{\mu}}x_{\mu,\vec{n}-\hat{\mu}}
\right)
=
0.
\label{eq:Gauss-law}
\end{eqnarray}
Here $\mu=1,2$ and $3$ stand for $x$, $y$ and $z$, respectively. 
Note that the coupling constant is written as $g_{\rm 1d}$ because this is the same as the coupling constant of the matrix model
(which can be regarded as `1d QFT'), which is used to obtain this action via the orbifold projection; see Appendix~\ref{sec:orbifold-projection} for details. The relation between this coupling and the coupling of the 4d theory will be shown shortly. 

Though we have introduced `lattice points' $\vec{n}$, 
it is not a legitimate lattice field theory yet; 
even the `lattice spacing' is missing, and there is no kinetic term!
Rather, at this moment, it should be called a quiver matrix model. 
To make a quiver matrix model into a lattice field theory, we use dimensional deconstruction \cite{ArkaniHamed:2001ca}. 
The basic idea of the dimensional deconstruction is that, when some fields in a quiver theory have nontrivial vacuum expectation values, a lattice structure for extra spatial dimensions can be generated effectively.  
In the case of the orbifold lattice, we assume that the fields $x, y$ and $z$ are fluctuating around $\frac{1}{\sqrt{2}ag_{\rm 1d}}\cdot\textbf{1}_k$, allowing us to identify a lattice spacing $a$ (we will see how to justify this assumption shortly).
Under this limit, writing $x, y, z$ and $A_t$ as
\begin{eqnarray}
x_{\vec{n}}
&=&
\frac{1}{\sqrt{2}ag_{\rm 1d}}\cdot\textbf{1}_k
+
\frac{a^{3/2}}{\sqrt{2}}\left(s_{1,\vec{n}}+iA_{1,\vec{n}}\right), 
\nonumber\\
y_{\vec{n}}
&=&
\frac{1}{\sqrt{2}ag_{\rm 1d}}\cdot\textbf{1}_k
+
\frac{a^{3/2}}{\sqrt{2}}\left(s_{2,\vec{n}}+iA_{2,\vec{n}}\right), 
\nonumber\\
z_{\vec{n}}
&=&
\frac{1}{\sqrt{2}ag_{\rm 1d}}\cdot\textbf{1}_k
+
\frac{a^{3/2}}{\sqrt{2}}\left(s_{3,\vec{n}}+iA_{3,\vec{n}}\right), 
\nonumber\\
A_{\vec{n}}
&=&
a^{3/2}A_{0,\vec{n}},
\label{eq:how-to-choose-background}
\end{eqnarray} 
where $s_{1,2,3}$ and $A_{1,2,3}$ are Hermitian, 
and\footnote{
The coupling constant $g^2_{\rm 4d}$ in \eqref{g_4d-vs-g_1d} is the bare coupling defined at the cutoff scale. 
} 
\begin{eqnarray}
g^2_{\rm 4d}=a^3g^2_{\rm 1d},
\label{g_4d-vs-g_1d}
\end{eqnarray} 
we obtain 
\begin{eqnarray}
L
=
\int d^3x{\rm Tr}
\left(
-
\frac{1}{4}F_{\mu\nu}^2
+
\frac{1}{2}(D_\mu s_I)^2
+
\frac{g_{\rm 4d}^2}{4}[s_I,s_J]^2
\right), 
\end{eqnarray}
up to $O(a)$ corrections. 
In this way, we can obtain a lattice regularization of $(3+1)$-d YM 
theory coupled to three scalar fields $s_{1,2,3}$. 

Actually, the form \eqref{eq:how-to-choose-background} 
and the requirement that $x,y,z\simeq\frac{1}{\sqrt{2}ag_{\rm 1d}}\cdot\textbf{1}_k$ are too restrictive. 
More precisely, 
we only need to have 
\begin{eqnarray}
x\bar{x}\simeq y\bar{y}\simeq z\bar{z}\simeq\frac{1}{2(ag_{1d})^2}\cdot\textbf{1}_k
\label{eq:how-to-choose-background-2}
\end{eqnarray}
to ensure the right continuum limit.
Here, $\simeq$ means the equality up to terms negligible as $a\to 0$. Note that such small fluctuations correspond to the scalar fields, as we will see shortly.

Note that the condition \eqref{eq:how-to-choose-background-2} is gauge-invariant, while $x,y,z\simeq\frac{1}{\sqrt{2}ag_{\rm 1d}}\cdot\textbf{1}_k$ is not.\footnote{
If $x_{\vec{n}}\simeq\frac{1}{\sqrt{2}ag_{\rm 1d}}\cdot\textbf{1}_k$, 
the gauge transformation \eqref{gauge-transf-scalar} maps such $x_{\vec{n}}$ to  
$\Omega_{\vec{n}}x_{\vec{n}}\Omega^{-1}_{\vec{n}+\hat{x}}\simeq \frac{1}{\sqrt{2}ag_{\rm 1d}}\cdot\Omega_{\vec{n}}\Omega^{-1}_{\vec{n}+\hat{x}}$. 
Hence, $x_{\vec{n}}\simeq\frac{1}{\sqrt{2}ag_{\rm 1d}}\cdot\textbf{1}_k$ is not a gauge-invariant statement. 
On the other hand, if $x_{\vec{n}}\bar{x}_{\vec{n}}\simeq\frac{1}{\sqrt{2}ag_{\rm 1d}}\cdot\textbf{1}_k$, 
it is transformed as $\Omega_{\vec{n}}x_{\vec{n}}\bar{x}_{\vec{n}}\Omega_{\vec{n}}^{-1}\simeq\frac{1}{\sqrt{2}ag_{\rm 1d}}\cdot\Omega_{\vec{n}}\Omega_{\vec{n}}^{-1}=\frac{1}{\sqrt{2}ag_{\rm 1d}}\cdot\textbf{1}_k$. 
Therefore, $x_{\vec{n}}\bar{x}_{\vec{n}}\simeq\frac{1}{\sqrt{2}ag_{\rm 1d}}\cdot\textbf{1}_k$ is a gauge-invariant statement. 
} The following parametrization \cite{Unsal:2005yh} gives us a convenient way to obtain the continuum theory in a gauge-invariant manner:
\begin{eqnarray}
x
&=&
\frac{1}{\sqrt{2}ag_{1d}}e^{a^{5/2}g_{\rm 1d}s_1}e^{ia^{5/2}g_{\rm 1d}A_1}, 
\nonumber\\
y
&=&
\frac{1}{\sqrt{2}ag_{1d}}e^{a^{5/2}g_{\rm 1d}s_2}e^{ia^{5/2}g_{\rm 1d}A_2}, 
\nonumber\\
z
&=&
\frac{1}{\sqrt{2}ag_{1d}}e^{a^{5/2}g_{\rm 1d}s_3}e^{ia^{5/2}g_{\rm 1d}A_3}.  
\nonumber\\
\label{relation-x-vs-U}
\end{eqnarray}
Clearly, the condition \eqref{eq:how-to-choose-background-2} is violated when the scalars $s_1,s_2$ and $s_3$
deviate too far from zero. This is known as the moduli stabilization problem \cite{Kaplan:2002wv}. Modulo this problem, 
the orbifold construction gives an alternative to more traditional lattice regularizations. 

We would now like to apply this construction to pure YM theory without the scalars $s_1,s_2$ and $s_3$. To accomplish this, we simply add
\begin{equation}
\Delta L_{\rm lattice}
\equiv
-\frac{m^2g_{\rm 4d}^2}{2a}\sum_{\vec{n}}
{\rm Tr}
\left( 
\left| x_{\vec{n}}\bar{x}_{\vec{n}} -\frac{1}{2a^2g_{\rm 1d}^2}\right|^2
+\left| y_{\vec{n}}\bar{y}_{\vec{n}} -\frac{1}{2a^2g_{\rm 1d}^2}\right|^2
+\left|z_{\vec{n}}\bar{z}_{\vec{n}} -\frac{1}{2a^2g_{\rm 1d}^2} \right|^2
\right),
\end{equation}
to the action in \eqref{eq:lattice-action}. 
In the continuum, this is nothing but the scalar mass term: 
\begin{eqnarray}
\Delta L
=
-\frac{m^2}{2}\int d^3x{\rm Tr}
\left(
s_1^2+s_2^2+s_3^2
\right). 
\end{eqnarray}
With such a mass term the expansion about $x\bar{x}\simeq y\bar{y}\simeq z\bar{z}\simeq\frac{1}{2(ag_{1d})^2}\cdot\textbf{1}_k$ is justified, 
resolving the moduli stabilization problem.
Furthermore, by taking $m^2$ large enough, we can completely eliminate the scalars. 
In the original references of the orbifold construction, 
this mass term could not be used because the main motivation was a supersymmetric lattice theory, and the scalar mass breaks supersymmetry. 
In our case, 
nothing forbids us from adding this scalar mass term 
since we are not interested in either supersymmetry or the scalar fields, for now. 
If, however, we consider the simulation of supersymmetric theories, 
then the moduli stabilization problem comes back. 
In this case
there are other ways to handle it; 
see Refs.~\cite{Hanada:2010qg,Hanada:2010kt,Hanada:2010gs,Hanada:2011qx} for resolutions. 

The biggest difference from the Wilson's plaquette action is the use of the non-compact variables. 
For the orbifold construction, the gauge-invariant path-integral measure is the flat measure on ${\mathbb R}^{2k^2}$, 
\begin{eqnarray}
\int dx_{\mu,\vec{n}}d\bar{x}_{\mu,\vec{n}}
=
\int_{-\infty}^\infty dx^{\rm (R)}_{\mu,\vec{n}}
\int_{-\infty}^\infty dx^{\rm (I)}_{\mu,\vec{n}},
\end{eqnarray}
where $x^{\rm (R)}_{\mu,\vec{n}}$ and $x^{\rm (I)}_{\mu,\vec{n}}$ are the real and imaginary parts of $x_{\mu,\vec{n}}$.
On the other hand, in the Wilson's plaquette action, the link variables are unitary, 
and the integral is defined using the Haar measure on the group manifold U($k$), which is compact. 
Note also that, because of \eqref{g_4d-vs-g_1d}, in the continuum limit ($a\to 0$, $g^{-2}_{4d}\sim -\log a$), 
the coupling constant $g_{1d}^2$ scales as $-\frac{1}{a^3\log a}$. Therefore, the continuum limit corresponds to the strong-coupling limit in the original matrix model.

In the orbifold construction, the U(1) part always appears by construction.\footnote{
Note that, even if the mother theory is taken to be SU($N$),
the traceless condition is not imposed on each block kept by the projection. 
} 
Hence it is impossible, or at least there is no known way, to construct the SU($k$) orbifold lattice.\footnote{
See, however, Ref.~\cite{Kanamori:2012et} for a proposal of the removal of the U($1$) part, for the Euclidean lattice. 
} 
However, it is not necessarily a problem. 
If all the fields are in the adjoint representation such as pure Yang-Mills or maximal super Yang-Mills, 
the U($1$) part is free and decoupled from SU($k$), 
so the local dynamics in the SU($k$) sector is not affected at all. 
It is not the case e.g., when matters in the fundamental representation are introduced, 
but still, the U($1$) sector can decouple from the low-energy dynamics if the U($1$) is not asymptotically free.\footnote{
The coupling constants for the SU($k$) and U($1$) parts are taken to be the same, asymptotically small value at the cutoff scale. From there, the SU($k$) coupling grows toward infrared, while the U($1$) coupling decreases.  
}

The generalizations to other dimensions are straightforward. For the $(2+1)$-dimensional theory, we use $\vec{n}=(n_x,n_y)$ to label the lattice points and the link variables $x_{\vec{n}}$ and $y_{\vec{n}}$. The lattice Lagrangian is obtained by omitting $z$ and $\bar{z}$ from \eqref{eq:lattice-action}. 
The coupling constant is $g_{\rm 3d}^2=a^2g_{\rm 1d}^2$. 
For the $(1+1)$-dimensional theory, we use $n=n_x$ as the label for the lattice points, and the link variable $x_n$. 
The lattice Lagrangian is obtained by omitting $y, \bar{y}, z$ and $\bar{z}$ from \eqref{eq:lattice-action}.
The coupling constant is $g_{\rm 2d}^2=ag_{\rm 1d}^2$. 
\subsection{Connection to the unitary-link formalism}\label{sec:KKU-vs-Wilson-Lagrangian}
\hspace{0.51cm}
Unlike the orbifold construction, the Kogut-Susskind formulation uses unitary link variables (see Appendix~\ref{sec:Kogut-Susskind} for a review.)
Using \eqref{relation-x-vs-U}, the connection between the KKU formalism and the formulation with unitary link variables can be seen rather straightforwardly, in the path-integral formulation. 
In the limit of infinite scalar mass $m\to\infty$, the scalars $s_1$, $s_2$ and $s_3$ are frozen to zero.
Then, due to \eqref{relation-x-vs-U}, 
the complex link variables $x, y$ and $z$ reduce to the unitary link variables up to a factor  $\frac{1}{\sqrt{2}ag_{1d}}$.  
The second line of \eqref{eq:lattice-action} becomes zero, and the third line becomes the magnetic plaquette term. 
A `Wilson loop' on the orbifold lattice, say ${\rm Tr}\left(x_{\vec{n}}y_{\vec{n}+\hat{x}}\bar{x}_{\vec{n}+\hat{y}}\bar{y}_{\vec{n}}\right)$, 
corresponds to a Wilson loop in the unitary-link formulation, say ${\rm Tr}\left(U_{x,\vec{n}}U_{y,\vec{n}+\hat{x}}U^\dagger_{x,\vec{n}+\hat{y}}U^\dagger_{y,\vec{n}}\right)$, up to an overall constant. When the scalar fields completely decouple, the flat measure for the non-compact variables reduces to the Haar measure for the compact ones. 
\subsection{Hamiltonian formalism (operator formalism)}\label{sec:operator-formalism}
\hspace{0.51cm}
The lattice Lagrangian obtained so far is nothing but a slightly complicated matrix model,
which is just the quantum mechanics of multiple variables. 
The path integral is defined using the flat measure in the same way as in a harmonic oscillator, 
or, more generally, a system of particles in flat space.\footnote{
In the unitary-link formulation, the path integral is evaluated with the Haar measure, 
making the commutation relations more intricate. See Sec.~\ref{sec:Kogut-Susskind} for details. 
} 
Hence, the Hamiltonian formulation can be obtained by the ordinary first-quantization procedure.\footnote{
Another way to obtain the same result is to quantize the matrix model and then perform the orbifold projection. 
} 
We use the standard $A_t=0$ gauge, forcing all physical states to be gauge-invariant because of the Gauss-law constraint \eqref{eq:Gauss-law}.\footnote{
Strictly speaking, some degrees of freedom cannot be gauged away. For example, when the time direction is compactified, the degrees of freedom corresponding to the Polyakov loop remain unfixed. The gauge-singlet constraint appears when these degrees of freedom are integrated out. For details, see e.g., Refs.~\cite{Hanada:2020uvt,Rinaldi:2021jbg}. }

The Hamiltonian can be written in terms of the link variables $x,y,z$, and their canonical conjugates $p_x,p_y,p_z$ as follows: 
\begin{eqnarray}
\hat{H}
&=&
\sum_{\vec{n}}
{\rm Tr}\Biggl(
|\hat{p}_{x,\vec{n}}|^2
+|\hat{p}_{y,\vec{n}}|^2
+|\hat{p}_{z,\vec{n}}|^2 \nonumber\\
&&\quad
+\frac{g_{\rm 1d}^2}{2}\left|
\hat{x}_{\vec{n}}\hat{\bar{x}}_{\vec{n}}
-\hat{\bar{x}}_{\vec{n}-\hat{x}}\hat{x}_{\vec{n}-\hat{x}}
+\hat{y}_{\vec{n}}\hat{\bar{y}}_{\vec{n}}
-\hat{\bar{y}}_{\vec{n}-\hat{y}}\hat{y}_{\vec{n}-\hat{y}}
+\hat{z}_{\vec{n}}\hat{\bar{z}}_{\vec{n}}
-\hat{\bar{z}}_{\vec{n}-\hat{z}}\hat{z}_{\vec{n}-\hat{z}} \right|^2 \nonumber\\
&&\quad
+2g_{\rm 1d}^2
\left( \left|
\hat{x}_{\vec{n}}\hat{y}_{\vec{n}+\hat{x}}
-\hat{y}_{\vec{n}}\hat{x}_{\vec{n}+\hat{y}} \right|^2
+\left|
\hat{y}_{\vec{n}}\hat{z}_{\vec{n}+\hat{y}}
-\hat{z}_{\vec{n}}\hat{y}_{\vec{n}+\hat{z}} \right|^2
+\left|
\hat{z}_{\vec{n}}\hat{x}_{\vec{n}+\hat{z}}
-\hat{x}_{\vec{n}}\hat{z}_{\vec{n}+\hat{x}} \right|^2 \right) \Biggl) \nonumber\\
&&\quad + \Delta\hat{H},
\label{KKU-Hamiltonian}
\end{eqnarray}
where
\begin{align}
\Delta\hat{H}
\equiv\frac{m^2g_{\rm 4d}^2}{2a}\sum_{\vec{n}}{\rm Tr}
\left(
\left|
\hat{x}_{\vec{n}}
\hat{\bar{x}}_{\vec{n}}
-
\frac{1}{2a^2g_{\rm 1d}^2}
\right|^2
+
\left|
\hat{y}_{\vec{n}}
\hat{\bar{y}}_{\vec{n}}
-
\frac{1}{2a^2g_{\rm 1d}^2}
\right|^2
+
\left|
\hat{z}_{\vec{n}}
\hat{\bar{z}}_{\vec{n}}
-
\frac{1}{2a^2g_{\rm 1d}^2}
\right|^2
\right).
\label{KKU-Hamiltonian-mass}
\end{align}

For simplicity we use $\hat{x}_\mu=\hat{x}, \hat{y}, \hat{z}$
and $\hat{p}_\mu=\hat{p}_x, \hat{p}_y, \hat{p}_z$
for $\mu=1,2,3$. Then the commutation relation can be written as
\begin{eqnarray}
[\hat{x}_{\mu\vec{n},pq},\hat{\bar{p}}_{\nu\vec{n}',rs}]=i\delta_{\mu\nu}\delta_{\vec{n}\vec{n}'}\delta_{ps}\delta_{qr},
\label{eq:commutation-relation-KKU}
\end{eqnarray}
and
\begin{eqnarray}
[\hat{x},\hat{p}]
=
[\hat{\bar{x}},\hat{\bar{p}}]
=
[\hat{x},\hat{x}]
=
[\hat{\bar{x}},\hat{\bar{x}}]
=
[\hat{p},\hat{p}]
=
[\hat{\bar{p}},\hat{\bar{p}}]
=
0. 
\end{eqnarray}

We are allowing $^{\rm (R)}$ and $^{\rm (I)}$ to denote real (Hermitian) and imaginary (anti-Hermitian) parts, such as\footnote{
The normalization factor $\frac{1}{\sqrt{2}}$ stems from the same factor in \eqref{complex-vs-Hermitian}, since we have chosen $\hat{x}^{\rm (R)}$ and $\hat{x}^{\rm (I)}$ to be independently normalized.}
  $\hat{x}=\frac{\hat{x}^{\rm (R)}+i\hat{x}^{\rm (I)}}{\sqrt{2}}$,
we obtain 
\begin{eqnarray}
[\hat{x}^{\rm (R)}_{\mu\vec{n},pq},\hat{p}^{\rm (R)}_{\nu\vec{n}',rs}]
=
[\hat{x}^{\rm (I)}_{\mu\vec{n},pq},\hat{p}^{\rm (I)}_{\nu\vec{n}',rs}]
=
i\delta_{\mu\nu}\delta_{\vec{n}\vec{n}'}\delta_{ps}\delta_{qr}. 
\label{eq:commutation-relation-KKU-Hermitian}
\end{eqnarray}
The Gauss-law constraint \eqref{eq:Gauss-law} becomes
\begin{eqnarray}
\hat{G}_{\vec{n},pq}
|{\rm phys}\rangle
=
0, 
\label{eq:gauge-invariant-state}
\end{eqnarray}
where
\begin{eqnarray}
\hat{G}_{\vec{n},pq}
\equiv
i\sum_{\mu=1}^3
\left(
-
\hat{x}_{\mu,\vec{n}}\hat{\bar{p}}_{\mu,\vec{n}}
+
\hat{p}_{\mu,\vec{n}}\hat{\bar{x}}_{\mu,\vec{n}}
-
\hat{\bar{x}}_{\mu,\vec{n}-\hat{\mu}}\hat{p}_{\mu,\vec{n}-\hat{\mu}}
+
\hat{\bar{p}}_{\mu,\vec{n}-\hat{\mu}}\hat{x}_{\mu,\vec{n}-\hat{\mu}}
\right)_{pq}. 
\end{eqnarray}
The operator $\hat{G}_{\vec{n}}$ generates local U($k$) gauge transformations. Indeed, 
\begin{align}
\left[
\sum_{r,s}
\epsilon^{rs}\hat{G}_{\vec{n},sr},\hat{x}_{\mu,\vec{n},pq}
\right]
=
-(\epsilon\hat{x}_{\mu,\vec{n}})_{pq}, 
\qquad
\left[
\sum_{r,s}
\epsilon^{rs}\hat{G}_{\vec{n},sr},\hat{x}_{\mu,\vec{n}-\hat{\mu},pq}
\right]
=
(\hat{x}_{\mu,\vec{n}-\hat{\mu}}\epsilon)_{pq}. 
\end{align}
Hence \eqref{eq:gauge-invariant-state} implies that physical states are gauge-invariant. 
\subsection{Symmetry at discretized level}
\label{sec:symmetry-orbifold-lattice}
\hspace{0.51cm}
We now summarize the symmetries present at the discretized level, before proceeding to regularize the Hilbert space in a way suitable for representation by qubits. 
\begin{itemize}
\item
The local U($k$) gauge symmetry. 

\item
Discrete translation. 

\item
Discrete rotation ($\frac{\pi}{2}$-degree rotation).
Note that the gauge field $A_{1,2,3}$ and scalar $s_{1,2,3}$ are rotated together. 

\item
Permutation of $x,y,z$. Note that the gauge field $A_{1,2,3}$ and scalar $s_{1,2,3}$ are permuted together.

\item
Charge conjugation $x,y,z\to x^\ast,y^\ast,z^\ast$, $A\to-A^\ast$. In the continuum, this is represented by $A_\mu\to -A_\mu^\ast$ and $s_I\to s_I^\ast$. 

\item
Parity symmetry $\vec{n}\to -\vec{n}$, 
$x_{\vec{n}},y_{\vec{n}},z_{\vec{n}}\to\bar{x}_{-\vec{n}-\hat{x}},\bar{y}_{-\vec{n}-\hat{y}},\bar{z}_{-\vec{n}-\hat{z}}$.

\end{itemize}
Notably, the orbifold lattice construction preserves all the same symmetries as the unitary-link formulation. 
This alone provides us with strong motivation to study the orbifold approach further in the context of quantum simulation. 

\section{Realization on a quantum computer}\label{sec:realization_on_QC}
\hspace{0.51cm}
In this section, we discuss a realization of the pure Yang-Mills theory on the quantum computer based on the orbifold construction.

As a concrete realization, we utilize the Fock basis. 
Alternatively, we could consider the coordinate and momentum bases, following the treatment in Refs.~\cite{Jordan:2011ne,Jordan:2011ci}.
We explain the latter in Appendix~\ref{sec:coordinate_basis}. 
\subsection{Fock space truncation}\label{sec:Fock-space-truncation}
\hspace{0.51cm}
The Hamiltonian \eqref{KKU-Hamiltonian} does not have a free-part. 
We introduce two parameters $\mu$ and $\omega$, and define `free part' just as a trick to introduce a Fock basis: 
\begin{eqnarray}
\hat{H}_{\rm free}
=
\sum_{\vec{n}}
{\rm Tr}\left(
\frac{
|\hat{p}_{x,\vec{n}}|^2
+
|\hat{p}_{y,\vec{n}}|^2
+
|\hat{p}_{z,\vec{n}}|^2}{\mu}
+
\mu\omega^2\left(
|\hat{x}_{\vec{n}}|^2
+
|\hat{y}_{\vec{n}}|^2
+
|\hat{z}_{\vec{n}}|^2
\right)
\right). 
\label{def:H_free}
\end{eqnarray}
Then, we write the Hamiltonian as
\begin{eqnarray}
\hat{H}
=
\hat{H}_{\rm free}
+
\hat{H}_{\rm int}. 
\end{eqnarray}
Note that, by definition, $\hat{H}_{\rm int}$ is chosen as $\hat{H}_{\rm int}\equiv \hat{H}-\hat{H}_{\rm free}$. 
The `mass' $\mu$ and `frequency' $\omega$ are free parameters that are used to split the Hamiltonian up into free and interacting parts. 
In general, the efficiency of the regularization may depend on $\mu$ and $\omega$. We define the creation and annihilation operators as 
\begin{eqnarray}
\hat{a}_{x,\vec{n}}^\dagger
=
\sqrt{\frac{\mu\omega}{2}}
\hat{x}_{\vec{n}}
-
\frac{i\hat{p}_{x,\vec{n}}}{\sqrt{2\mu\omega}}, 
\qquad
\hat{a}_{x,\vec{n}}
=
\sqrt{\frac{\mu\omega}{2}}
\hat{x}_{\vec{n}}
+
\frac{i\hat{p}_{x,\vec{n}}}{\sqrt{2\mu\omega}}, 
\end{eqnarray}
and similarly for $y$ and $z$. Each of $x_{\vec{n}}$, $y_{\vec{n}}$ and $z_{\vec{n}}$ is a $k\times k$ complex matrix having $2k^2$ real degrees of freedom, while $\vec{n}$ labels $L^3$ different lattice points.  
Hence the number of harmonic oscillators describing the Fock space is $3\times 2k^2\times L^3$. 
We truncate the Hilbert space such that the excitation level of each oscillator is below $\Lambda$.
The dimension of this truncated Hilbert space is $\Lambda^{6k^2L^3}$. 
The lattice theory is reproduced when this cutoff is removed,\footnote{
Here, we have implicitly assumed that the states under consideration are bounded both in coordinate and momentum spaces. Otherwise, this cutoff procedure may not make sense; for example, to approximate a state $\ket{x}=\int dp e^{ipx}\ket{p}$, infinitely large excitation modes are needed. This assumption is valid as long as the energy density is finite. 
} i.e.,~$\Lambda\to\infty$.

Next we introduce an explicit expression in terms of qubits. We use the prescription used in Ref.~\cite{matrixModel}. 
Let $|j\rangle$ ($j=0,1,\cdots,\Lambda-1$) be the $j$-th excited state of the harmonic oscillator. 
We can write $j$ in terms of binaries as 
$j = \sum_{l=0}^{K-1}b_l2^l$. 
By using $K\equiv\log_2\Lambda$ qubits, we can rewrite the state $\ket{j}$ as 
\begin{align}
|j\rangle  = \left| {{b_{0}}} \right\rangle \left| {{b_{1}}} \right\rangle  \ldots \left| {{b_{K-1}}} \right\rangle ~.
\end{align}

With this encoding, the creation operator takes the form
\begin{align}
\hat{a}^\dag = \sum\limits_{j = 0}^{\Lambda - 2} {\sqrt {j + 1} } |j + 1\rangle \langle j|. 
\label{adag-in-Fock-basis}
\end{align}
Writing $|j\rangle  = \left| {{b_{0}}} \right\rangle \left| {{b_{1}}} \right\rangle  \ldots \left| {{b_{K-1}}} \right\rangle$
and $|j+1\rangle  = \left| {{b'_0}} \right\rangle \left| {{b'_1}} \right\rangle  \ldots \left| {{b'_{K-1}}} \right\rangle$, 
we can express $|j + 1\rangle \langle j|$ as an operator in this basis as
\begin{align}
 |j + 1\rangle \langle j|
 = \otimes_{l=0}^{K-1}
 \left(|b'_l\rangle\langle b_l|\right). 
 \label{adag-in-Fock-basis-2}
\end{align}
Note that each $|b'_l\rangle\langle b_l|$ is a linear combination of the Pauli matrices: 
\begin{eqnarray}
& &
|0\rangle\langle 0|=\frac{\textbf{1}_2-\sigma_z}{2}, 
\qquad
|1\rangle\langle 1|=\frac{\textbf{1}_2+\sigma_z}{2}, 
\nonumber\\
& &
|0\rangle\langle 1|=\frac{\sigma_x+i\sigma_y}{2}, 
\qquad
|1\rangle\langle 0|=\frac{\sigma_x-i\sigma_y}{2}.  
 \label{adag-in-Fock-basis-3}
\end{eqnarray}
Therefore, $\hat{a}^\dagger$ can be written as a linear combination of 
Pauli strings of length $K=\log_2\Lambda$ (i.e. a tensor product of $K$ Pauli spin operators). 
The same holds for $\hat{a}$, and hence, 
$\hat{x}_{\vec{n}}$, $\hat{y}_{\vec{n}}$ and $\hat{z}_{\vec{n}}$ are linear combinations of such Pauli strings.
Each creation or annihilation operator consists of less than $\Lambda^2$ Pauli strings,\footnote{
Each $\ket{j+1}\bra{j}$ in \eqref{adag-in-Fock-basis} is written as 
a sum of less than $\Lambda=2^K$ Pauli strings, because each $\ket{b'_l}\bra{b_l}$ in 
\eqref{adag-in-Fock-basis-2} contains one or two Pauli matrices. 
} 
so each four-point interaction contains at most $\Lambda^8$ Pauli strings. There are $O(k^4)$ number of combinations regarding the color indices\footnote{
For example, the plaquette ${\rm Tr}(xy\bar{x}\bar{y})$ can be written as 
$\sum_{a,b,c,d=1}^kx_{ab}y_{bc}\bar{x}_{cd}\bar{y}_{da}$, and hence there are $k^4$ combinations of $(a,b,c,d)$. 
},
and we must multiply this by the lattice volume (number of lattice sites) $L^3$ to obtain the total number of Pauli strings. 
Thus, the number of Pauli strings is bounded above by $L^3\Lambda^8k^4$, up to a numerical constant, where each Pauli string is of length $4K=4\log_2\Lambda$ at most.

The free part $\hat{a}^\dagger\hat{a}$ can also be expressed using Pauli strings, but the cost of the free part is negligible compared to the interaction part, so we do not consider it here.
\subsection{Gauge-singlet constraint}\label{sec:singlet-constraint}
\hspace{0.51cm}
By sending the cutoff $\Lambda$ to $\infty$, we obtain the lattice Hamiltonian acting on the extended Hilbert space containing the gauge non-singlet states. 
As with the Kogut-Susskind formulation, in general, it is difficult to truncate this extended Hilbert space directly to the subspace of physical gauge-invariant states. Therefore, we must choose the initial state to be a gauge singlet and simulate time-evolution precisely enough for the state to remain gauge invariant. Alternately, by adding a term like $\sum_{\vec{n}}{\rm Tr}\hat{G}^2_{\vec{n}}$ to the Hamiltonian, we can penalize the violation of the gauge-singlet constraint so that the gauge-singlet constraint is maintained in low-energy processes.
\subsection{Ground state preparation}\label{sec:ground-state-preparation}
\hspace{0.51cm}
In the Lagrangian formulation, we imposed the condition \eqref{eq:how-to-choose-background-2} to obtain the desirable continuum limit. As the counterpart of this condition in the operator formulation,
the gauge-invariant ground state $|{\rm VAC}\rangle$ (which is {\it not} the Fock vacuum) satisfies
\begin{eqnarray}
\left(\hat{x}_{\vec{n}}\hat{\bar{x}}_{\vec{n}}\right)_{pq}
|{\rm VAC}\rangle
\simeq
\left(\hat{y}_{\vec{n}}\hat{\bar{y}}_{\vec{n}}\right)_{pq}
|{\rm VAC}\rangle
\simeq
\left(\hat{z}_{\vec{n}}\hat{\bar{z}}_{\vec{n}}\right)_{pq}
|{\rm VAC}\rangle
\simeq
\frac{\delta_{pq}}{2a^2g_{\rm 1d}^2}|{\rm VAC}\rangle.
\end{eqnarray}
Here $\simeq$ indicates the equality up to terms that disappears in the continuum limit $a\to 0$, as in \eqref{eq:how-to-choose-background-2}.
This does not imply, however, that $\hat{x}_{\vec{n},pq}|{\rm VAC}\rangle\simeq\frac{\delta_{pq}}{\sqrt{2}ag_{\rm 1d}}|{\rm VAC}\rangle$, as such a condition is not gauge invariant. To construct this ground state we will apply the adiabatic algorithm. The approach here will be almost identical to the construction discussed in Section 4.5 of \cite{matrixModel}, where block-encoding of the Hamiltonian is achieved from the Pauli sum form and then Wan-Kim \cite{Wan2020FastDM} algorithm is carefully applied to efficiently prepare the ground state. \footnote{In this article we will not go into details, but recommend the reader to consider reading Ref.~\cite{matrixModel}}
For $0\le s\le 1$, we introduce $\hat{H}(s)$ as 
\begin{eqnarray}
\hat{H}(s)
=
(1-s)\hat{H}_{\rm free}
+
s\hat{H}. 
\end{eqnarray}
At $s=0$ the Hamiltonian is $H_{\rm free}$ defined by \eqref{def:H_free}, and
we can simply choose the Fock vacuum $|0\rangle$ as the gauge-invariant ground state. 
Then, we gradually change $s$ from 0 to 1 to prepare $|{\rm VAC}\rangle$. The quantum gate complexity of ground state preparation algorithm is analogous to analysis in Ref.~\cite{matrixModel} and given by
$$O\left(\frac{C \beta^2}{\Delta_{\text{gap}}^2} \text{polylog} \left( \frac{\beta }{\Delta_{\text{gap}}} \frac{1}{\delta}\right) \right), $$
where $C\sim L^3k^4\Lambda^{8}\log_2\Lambda$ and $\beta \sim g^2k^4L^3\Lambda^6$ in our setup and $\delta$ is the error (in 1-norm distance) of constructed state from the true ground state. 
Note that this complexity is controlled by the mass gap $(\Delta_{\text{gap}})$ of the adiabatic Hamiltonian $H(s)$. 
Note also that, in the current setup, 
it is better to know the behavior of the gap in the extended Hilbert space, not just in the gauge-singlet sector, 
since the time evolution cannot be perfectly gauge-invariant due to various errors.\footnote{
For example, the truncation of the Fock space breaks the gauge invariance. Also, depending on the detail of the algorithm, the gauge invariance may be broken small amount. 
As mentioned before, it might be possible to avoid this issue simply by adding a term proportional to $\sum_{\vec{N}}{\rm Tr}\hat{G}^2$.
} 
One way to estimate the gap is to calculate 
the expectation value of the energy 
as a function of temperature $T$, by using Monte Carlo simulation of the un-gauged Euclidean theory, in which the gauge field $A_t$ is turned off.
At low temperature, 
the energy should approach the ground state value $E_0$ 
as 
$E(T) = E_0+ (E_0 +\Delta E )e^{-\Delta E/T} +{O}\left( e^{-(E_2 -E_0)/T} ,e^{-2\Delta E/T} \right)$, 
where $\Delta E$ is the energy gap. 
(Note that this calculation gives the gap at $\Lambda=\infty$.)
Such analysis has already been done for the matrix model (mother theory) at $s=1$ \cite{Berkowitz:2018qhn,Maldacena:2018vsr}. We defer detailed consideration of non-adiabatic errors to future work.

\subsection{Optimal choice of regularization parameter $\mu$ and $\omega$}\label{sec:choice-of-free-parameters}
\hspace{0.51cm}
The mass $\mu$ and frequency $\omega$ in $H_{\rm free}$ are parameters associated with the regularization of the Hilbert space. 
Depending on the choice of $\mu$ and $\omega$, the finite-cutoff effect behaves differently as the cutoff is removed. 
What would be the optimal choice of $\mu$ and $\omega$, which leads to efficient truncation? 

In the orbifold construction, the continuum theory is described by the fluctuations about the background \eqref{eq:how-to-choose-background-2}. We use the Fock states to describe this background and the fluctuations. 
Therefore, the wave functions of the low-lying Fock states have to be not-too-large and not-too-small, such that 
the background \eqref{eq:how-to-choose-background-2} is described efficiently. 
Because typical sizes of the wave function is given by $\bra{n}\hat{x}^2\ket{n}=\frac{n+\frac{1}{2}}{\mu\omega}$ for the $n$-th excited state,
$\mu\omega\sim (ag_{\rm 1d})^2$ is a natural choice. 
Then, low-lying (small-$n$) modes form the background efficiently, 
and large-$n$ modes describe high-frequency fluctuations.
\subsection{Example of efficient time-evolution algorithm}
\hspace{0.51cm}
One apparent advantage of the KKU formulation is the simplicity of the Hamiltonian in the Fock basis. 
As explained in Sec.~\ref{sec:Fock-space-truncation}, it takes the form
\begin{eqnarray}
\hat{H}
=
\sum_{i=1}^{n_{\rm P.s.}}\alpha_i \hat{S}_i, 
\qquad
n_{\rm P.s.}\lesssim L^3\Lambda^8k^4,
\end{eqnarray}
where $\hat{S}_i$ are Pauli strings of length $4\log_2\Lambda$ at most. 
The Pauli strings are unitary operators, which can easily be expressed in terms of basic quantum gates. 

Again, this form of the Hamiltonian is essentially the same as the one used for our matrix model paper in Ref.~\cite{matrixModel}. Therefore, the same sort of algorithms can be used for efficient quantum simulations as described in Section 4.4 of Ref.~\cite{matrixModel}. More specifically, one can perform block-encoding and qubitization ~\cite{low2016hamiltonian} of the Hamiltonian (in Pauli sum form) and then apply Quantum Signal Processing (QSP)~\cite{low2017optimal} approach.\footnote{Again we will not go into details here, since interested reader can read the details in Ref.~\cite{matrixModel}.}  
By definition, time evolution is described by the unitary operator $e^{-i\hat{H}t}$. 
The QSP uses the Jacobi-Anger expansion of the time evolution operator   
\begin{equation}
e^{-i\hat{H}t} = J_0(-\lambda t) + 2 \sum_{n=1}^\infty i^n  J_{n}(-\lambda t) \times T_n\Big(\frac{\hat{H}}{\lambda}\Big), 
\label{Jacobi-Anger}
\end{equation}
where $J_n$ is the Bessel function of the first kind and 
$T_n$ is the Chebyshev polynomial of the first kind, providing us with an efficient way to implement this expansion on a digital quantum computer. 
As explained in Ref.~\cite{matrixModel}, the QSP, combined with 
the treatment of qubitization in Ref.~\cite{low2016hamiltonian}, provides us with an efficient implementation of the right-hand side of \eqref{Jacobi-Anger}. 

In the implementation of the QSP, the operators $\hat{R}$ and $\hat{U}$ constructed in Ref.~\cite{matrixModel} are multiplied to the quantum state repeatedly. 
In order to approximate $e^{-i\hat{H}t}$ up to error $\epsilon$, 
the necessary number of applications of $\hat{R}$ and $\hat{U}$ is 
$O\left(C \left(||\alpha||\cdot t+\log\epsilon^{-1}\right)\right)$
\cite{Babbush:2018mlj}, where $C$ is the cost of multiplying $\hat{U}$ and $\hat{R}$ to a quantum state and $||\alpha||=\sum_{i=1}^{n_{\rm P.s.}}|\alpha_i|$. 
In our setup, $C\sim L^3k^4\Lambda^{8}\log_2\Lambda$ and $||\alpha||\sim g_{\rm 1d}^2k^4L^3\Lambda^6$ in our setup.
Therefore, the increase of the cost is much slower compared to the growth of the dimension of the Hilbert space, $\Lambda^{6k^2L^3}$.
\subsection{Measuring glueballs}
\hspace{0.51cm}
In order to perform useful computations, it is not enough to prepare a state; we must also be able to measure interesting observables.  In pure Yang-Mills theory, for instance, correlation functions of glueball operators in the confining vacuum are a subject of intense interest where quantum computation may offer a distinct advantage \cite{Glueball1999,doi:10.1063/1.1843694,yamanaka2019glueball}. In addition, measuring the occupation numbers of glueball modes is important for computing amplitudes in gauge theory scattering processes, an exciting prospect for quantum simulation. These glueball operators are constructed as linear combinations of the Wilson loops, which are gauge-invariant, path-ordered products of unitary link variables, where the path is chosen according to the operator's representations under charge conjugation, parity, and the point group associated with the lattice.

As discussed around eq.~\eqref{relation-x-vs-U}, when the scalar fields $s_{1,2,3}$ are sufficiently suppressed by a large mass, we have\footnote{
If the scalar fields are not suppressed, the loop obtained from $\hat{x}_{\vec{n},\mu}$ resembles the supersymmetric Wilson loop, which is frequently considered in the context of gauge/gravity duality \cite{Maldacena:1998im}. 
}
\begin{equation}
\hat{x}_\mu \simeq \frac{1}{\sqrt{2}ag_{1d}}\hat{U}_\mu, 
\end{equation}
where $U_i$ is the unitary link variable. 
Thus, for a glueball operator $\Phi^R$ composed of closed paths $\gamma$ of length at most $l$,
\begin{equation}\label{eq:glueball_operators}
\hat{\Phi}^R = 
\sum_{\gamma} \tilde{c}_{\gamma}{\rm Tr} \left(\prod\limits_{\{\vec{n},\mu\}_{\gamma}} \hat{x}_{\vec{n},\mu} \right)
\simeq
\sum_{\gamma} c_{\gamma}{\rm Tr}\left( \prod\limits_{\{\vec{n},\mu\}_{\gamma}} \hat{U}_{\vec{n},\mu}\right).
\end{equation}

In this way, the problem of measuring correlation functions of glueball operators is reduced to computing expectation values of products of the scalar fields $\prod x_{\vec{n},\mu}$ along closed loops. Note that, in this formulation, the link variables $x_{\vec{n},\mu}$ do not act on the Hilbert space of group elements of $\text{U}(k)$, but are rather embedded in the larger space ${\mathbb C}^{k^2}$, truncated appropriately to our regularization scheme.

The canonical method of measuring observables is to use the phase estimation technique introduced by Kitaev~\cite{Kitaev:1995qy}. This allows one to directly compute the expectation value of $\Phi^R$ to precision $\varepsilon$ using at most ${O}(\text{log}(1/\varepsilon))$ ancilla qubits and ${O}(1/\varepsilon)$ controlled-$U$ operations, where $U=e^{i\Phi^R}$. This protocol works by applying the unitary $U$ to the desired state $\ket{\psi}$ a number of times depending on the state of a register of $l$ qubits, where $l={O}(\text{log}(1/\varepsilon))$, applying the inverse Fourier transform to the register, and conducting a measurement on the register \cite{Lloyd}. Repeating this procedure many times, one can obtain an estimate of the expectation value of the desired operator.

Alternatively, one can use the algorithm developed recently by Huang, Kueng, and Preskill~\cite{Huang2020} implementing shadow tomography to efficiently predict expectation values of local observables. This procedure carries several advantages over phase estimation; first, the algorithm uses a number of gates only polylogarithmic in the number of observables to be measured. As long as the state to be measured can be prepared efficiently, this represents an exponential speedup relative to phase estimation. Second, the protocol is well-suited for near-term quantum devices, avoiding the need for long coherence times, ancilla qubits, and controlled operations. 

The algorithm predicts the expectation values of $M$ local observables using the following method:
\begin{itemize}
\item[1.] Apply a random single-qubit Clifford circuit\footnote{The Clifford group $\mathcal{C}_n$ is the normalizer of the Pauli group on $n$ qubits $G_n$, i.e.,~the set of elements $g$ which satisfy $gG_n=G_ng$.  
Here, a random single-qubit Clifford circuit means a tensor product of elements of $C_1$ over each qubit. This is equivalent to measuring each qubit in a random Pauli basis.}, then measure each qubit in the computational basis. Repeat this step $N = {O}(\text{log}(M))$ times.
\item[2.] Use the measurement outcomes to construct $N$ classical representations of the state using the technique of classical shadows (this can be done efficiently when the observables $\{O_1,...,O_M\}$ are local, see Ref.~\cite{Huang2020}), then compute the expectation value of each observable with respect to these classical representations.
\item[3.] Group the outcomes into $m$ equal-sized groups, and compute the means of all observables for each group. For each of the $M$ observables, the prediction for its expectation value is the median of these means.
\end{itemize}
For $M$ observables $\{ O_1,...,O_M\}$ each acting on at most $r$ qubits, this suffices to predict $\{\langle O_1 \rangle,...,\langle O_M\rangle\}$ to precision $\varepsilon$ with probability $1-\delta$, saturating information theoretic lower bounds, where
\begin{eqnarray}
N &=& (2\text{log}(2M/\delta))\frac{34}{\varepsilon^2}4^{r} \text{max}_i||O_i||_{\infty}^2~, \nonumber\\
m &=& 2\text{log}(2M/\delta)~.
\end{eqnarray}

The glueball operators are composed out of Wilson loops as in Eq.~\eqref{eq:glueball_operators}. We treat the maximal length of these Wilson loops and the number of Wilson loops per glueball operator as unknown constants. In this case, the number of qubits a loop acts on is of order $K k^2 = k^2\log_2\Lambda$. Let $f(\psi)$ be the time it takes to prepare the state $\ket{\psi}$ to be measured. Then, the shadow tomography protocol suffices to measure all the specified glueball operators, $\{\Phi^R_i\}$, to precision $\varepsilon$ with probability $1-\delta$ in time $O\left(4^{K k^2}\text{max}_i ||\Phi^R_i||_{\infty} \ \text{log}(L^3/\delta)f(\psi)/\varepsilon^2\right)$.
\subsection{Other observables}
\hspace{0.51cm}
Measuring glueball operators is clearly an interesting task, but we can envision numerous use cases for quantum computing where other observables are relevant. For instance, to investigate topological physics, one could introduce a theta term into the Hamiltonian and study the topological charge over the extent of a spatial lattice. For applications in nuclear physics, one may be interested in the spatial distribution of the energy or action density. These applications demand knowledge of the expectation values of a very large number of local observables, making shadow tomography especially practical. Any physical (gauge-invariant) observable can be constructed out of Wilson loops, similar to the construction of glueball operators above, so the same techniques used for measuring their expectation values may be applied. 

\subsection{Jordan-Lee-Preskill bound on Hilbert space}
\hspace{0.51cm}
In this section, following the discussion in a series of papers by Jordan, Lee, and Preskill (JLP) \cite{Jordan:2011ne,Jordan:2011ci}, we consider how to upper-bound the probability of a state being outside the truncated Hilbert space as a function of its energy. Intuitively, this provides a justification for imposing a cutoff $\Lambda$ on the local dimension of the Hilbert space, provided we do not probe physics near or above a given energy scale $E(\Lambda)$. 

There are two main differences between our approach and that taken by JLP. First, we truncate with respect to the Fock basis rather than the coordinate basis used by JLP. 
Second, in JLP, one not only has to truncate the field range, but also the number of discrete values the field is allowed to take, at least for bosonic lattice field theories. In our case, we truncate the maximal occupation number on each bosonic mode. The same arguments go through in each case with minimal modification, however.
See Appendix~\ref{sec:coordinate_basis} regarding more about the regularization in the coordinate basis. 

Specifically, we truncate each oscillator at occupation number $\Lambda$, so that we require $6k^2\text{log}_2\Lambda$ qubits per site. We wish to show that we can simulate any physical process below an energy scale $E$ up to error $\varepsilon$ with $\Lambda$ at most polynomial in $1/a,1/\varepsilon,L^3$. Let $P_{\Lambda}$ be the projection operator onto the subspace where no oscillator has occupation number greater than or equal to $\Lambda$. Then, we define
\begin{eqnarray}
\bra{\psi}P_{\Lambda}\ket{\psi} &\equiv& 1-p_{\text{out}} \nonumber\\
&\geq& 1-6k^2L^3\text{max}(p_{\text{out}}(\vec{n},ij,\nu,\sigma)) \nonumber\\
&\equiv& 1-6k^2L^3\text{max}(p_{\text{out}}(\mathbf{x})).
\end{eqnarray}
Here, $\vec{n}$ labels the spatial sites, $i,j$ label the matrix elements, $\nu$ labels the spatial direction, and $\sigma$ labels the real and imaginary parts. The notation $\mathbf{x}$ is simply a shorthand for all of these indices. The name $p_{\text{out}}(\mathbf{x})$ refers to the probability that oscillator $\mathbf{x}$ is found outside the truncated Hilbert space.

Let $\mu_{n(\mathbf{x})}$ and $\sigma_{n(\mathbf{x})}$ be the mean and standard deviation, respectively, of the occuptaion number of the oscillator labeled by $\mathbf{x}$. By Chebyshev's inequality, if $\Lambda = |\mu_{n(\mathbf{x})}|+c\sigma_{n(\mathbf{x})}$, $c > 0$, then
\begin{equation}
p_{\text{out}} \leq \frac{1}{c^2}.
\end{equation}
Thus, by choosing
\begin{equation}
\Lambda = \text{max} \left( |\mu_{n(\mathbf{x})}|+\sqrt{\frac{6k^2L^3}{\varepsilon}}\sigma_{n(\mathbf{x})} \right),
\end{equation}
we have $\bra{\psi}P_{\Lambda}\ket{\psi} \geq 1-\varepsilon$.

By definition,
\begin{eqnarray}
\mu_{n(\mathbf{x})} &=& \bra{\psi}\hat{n}_{\mathbf{x}}\ket{\psi}~, \nonumber\\
\sigma_{n(\mathbf{x})} &=& \sqrt{\bra{\psi}\hat{n}^2_{\mathbf{x}}\ket{\psi} - \bra{\psi}\hat{n}_{\mathbf{x}}\ket{\psi}^2}~.
\end{eqnarray}
Since $|\bra{\psi}M\ket{\psi}| \leq \sqrt{\bra{\psi}M^2\ket{\psi}}$, we have that
\begin{equation}
\Lambda = O\left( \sqrt{\frac{k^2L^3}{\varepsilon}\max\limits_{\mathbf{x}} \bra{\psi}\hat{n}^2_{\mathbf{x}}\ket{\psi}} \right)~.
\end{equation}
All we need to do, then, is upper bound $\max_{\mathbf{x}}\bra{\psi}\hat{n}^2_{\mathbf{x}}\ket{\psi}$ as a function of the energy $E = \bra{\psi}\hat{H}\ket{\psi}$. This is particularly easy for $\hat{H}_{\text{free}}$, which is a sum of $6k^2L^3$ independent oscillators of mass $\mu$ and frequency $\omega$. In that case, we overestimate the maximum of $\text{max}_{\mathbf{x}}\bra{\psi}\hat{n}_{\mathbf{x}}^2\ket{\psi}$ over all states $\ket{\psi}$ with energy $E$ by putting one oscillator in the $\ket{n}$ state and all others in the $\ket{0}$ state, where $n = \lceil \frac{E-\omega/2}{\omega} \rceil$ (i.e., the smallest integer which satisfies $n\ge\frac{E-\omega/2}{\omega}$). Then, we find for the free case $\Lambda = O\left( \frac{kL^{3/2}E}{\sqrt{\varepsilon} \omega} \right)$.

Unfortunately, in the interacting theory we find no simple analytic bounds on $\max_{\mathbf{x}}\bra{\psi}\hat{n}^2_{\mathbf{x}}\ket{\psi}$ as a function of $E$. However, one could imagine testing this numerically on a classical or quantum computer via a well-controlled procedure. Estimating this error is crucial to understanding the resource requirements of simulating high energy scattering processes. Bounds of the type found by JLP are likely to be quite loose in practice, so there is an independent motivation for studying these errors numerically in the case of bosonic lattice field theory.

Alternatively, one could consider truncating our Hamiltonian using the coordinate and momentum basis instead of the Fock basis. In this case, it may be possible to derive upper bounds on the truncation error by bounding the fluctuations of coordinates and momenta. For an interacting theory, this bound would potentially be quite weak. In general, in order to claim stronger bounds, additional physical ingredients are needed, for instance, specifying which terms of the full Hamiltonian contribute the most energy to some physically well-motivated state. Some related comments on the efficiency of the choice of basis are given in \cite{Klco:2018zqz}.

\section{From orbifolds to Kogut-Susskind}\label{sec:comparison}
\hspace{0.51cm}
Both the orbifold construction (KKU) and the Kogut-Susskind formulation (KS) have pros and cons. 
Let us compare several aspects: 

\begin{itemize}
\item
The KS requires complicated group theory associated with the harmonic expansion on the group manifold, while the KKU requires almost no group theory, using the Fock basis of ordinary harmonic oscillators instead. 

This simplification in the link variables facilitates representing the Hilbert space with qubits and may open the door to the quantum simulation of lattice gauge theories with continuous variables, where harmonic oscillators are used routinely \cite{continuous2005}.

\item
In the KKU, the Hamiltonian is a linear combination of Pauli strings, which is easy to handle. A method of transcribing the KS formulation into Pauli operators was given by Byrnes and Yamamoto \cite{PhysRevA.73.022328}. However, their approach requires the calculation of Clebsch-Gordan coefficients for U($N$) and a significant amount of classical pre-processing\footnote{It is not trivial to determine Clebsch-Gordan coefficients. However, algebraic methods are known for SU($N$) (for small $N$) and U($N$), and a numerical method is known for SU($N$) for all $N$ \cite{alex2011numerical,Rowe:Algorithm,biedenharn1968}. Quantum algorithms are also known, which appear to give speedups over classical algorithms \cite{PhysRevLett.97.170502,jordan2008fast}.}. As a result, the form of the Pauli strings composing the U($N$) KS Hamiltonian is unclear a priori.

\item
No simple truncation to an orthonormal basis of the gauge-invariant Hilbert space is known in either formulation. 

\item
The level of complication regarding the preparation of the quantum vacuum of the interacting theory appears to be the same.
There may be a large difference in the efficiency of the two approaches, though, particularly with regard to the amount of truncation required for certain precision.

\item
In the KKU, it is straightforward to construct the U($k$) theory, and it may be possible to realize SU($k$), O($k$), and Sp($k$) theories in a similar manner.
The same holds for the KS, with the reservation that the representation theory may be complicated for generic gauge groups.  

\item
The KKU is subject to the moduli stabilization problem, while KS does not have this problem. Note, however, that in many cases, including pure Yang-Mills, the moduli stabilization problem can be resolved. 

\item
In the KKU, some supersymmetric theories can be realized without parameter fine tuning~\cite{Kaplan:2002wv}. This may or may not be possible in the KS.\footnote{
Perhaps the Hamiltonian version of Sugino's lattice action~\cite{Sugino:2003yb,Sugino:2004qd,Sugino:2004uv} can be constructed. }

\item
The KKU may require more qubits because half of the degrees of freedom are dynamically eliminated associated with the moduli fixing. 
It requires careful analysis to see if this is actually the case, though. 

\end{itemize}

Our conclusion is that the KKU and KS form complementary approaches. There may be problems or devices for which the orbifold approach is more natural.

At the level of the path integral, the connection between the KKU model and the unitary-link formulation
is very simple, as explained in Sec.~\ref{sec:KKU-vs-Wilson-Lagrangian}. 
As we can see from \eqref{relation-x-vs-U}, the complex link variables $x$, $y$ and $z$ reduce to the unitary link variables ($\sim$angular components)
when the scalars (radial components) decouple. 
In the operator formalism, the difference between the commutation relations \eqref{eq:commutation-relation-KKU}
and \eqref{eq:commutation-relation-KS} may cause some nontrivial deviation at finite lattice spacing.

As we have seen in eq.~\eqref{g_4d-vs-g_1d} in Sec.~\ref{sec:orbifold-lattice-explicit-construction}, 
the parameter $g_{\rm 1d}^2$ in the orbifold lattice is related to the lattice spacing $a$ and coupling constant $g_{\rm 4d}^2$ in the 4d theory as 
$g_{\rm 1d}^2=a^{-3}g_{\rm 4d}^2$. 
Therefore, the coarse-lattice limit $a\to\infty$ is the weak-coupling limit $g_{\rm 1d}^2\to 0$. 
From \eqref{KKU-Hamiltonian}, we can see that only the ${\rm Tr}|\hat{p}|^2$ term survives there.
If we add a large scalar mass as in \eqref{KKU-Hamiltonian-mass}, the scalar part can be decoupled. 
Including this term, the weak-coupling limit of the orbifold lattice Hamiltonian is analogous to the strong-coupling limit of the KS formulation.

The truncation of the higher-excited modes in Fock space naturally restricts the momentum, 
because $\bra{n}\hat{p}^2\ket{n}$ grows linearly with $n$. This resembles the cutoff of the electric field (equivalently, the cutoff for the size of the representation) in the KS formulation.
Both schemes naturally discard the high-energy modes in the weak-coupling limit of the KKU or the strong-coupling limit of the KS.

Finally, we mention briefly that other important reformulations of lattice gauge theory exist, notably the quantum link model approach (see the reference \cite{CHANDRASEKHARAN1997455}). This kind of model might be practical when performing Monte Carlo calculations on a classical computer, where the classical lattice gauge theory action is replaced with a quantum mechanical counterpart. This is potentially useful when performing the Monte Carlo path integral in the context of quantum computation (see \cite{Lamm:2019bik}).  


\section{Conclusion and Outlook}\label{sec:conclusion}
\hspace{0.51cm}
In this paper, we have demonstrated how to apply the orbifold methods used by KKU \cite{Kaplan:2002wv}
for constructing supersymmetric lattice gauge theories to the quantum simulation of ordinary Yang-Mills theory with U($k$) gauge group. 
In particular, we have focused on several prototypical tasks concerning the quantum simulation of high-energy processes, including preparation of the interacting vacuum state and measurement of local dynamic observables. Our construction reframes the problem of simulating the Yang-Mills theory as one of simulating a large collection of coupled harmonic oscillators. 
We considered the $(3+1)$-dimensional Yang-Mills theory for concreteness, although the generalizations to other dimensions are straightforward. 
We are hopeful that this novel approach will enable further exploration into the advantages of quantum computing in simulating elementary particle physics and serve as a practical formalism for some experimental quantum computing platforms. We defer a detailed comparison of the efficiency of various approaches for specific computational problems to future work and elaborate on a number of related open problems below.
\subsection{Hamiltonian formulation and quantum simulation}
\hspace{0.51cm}
We wish to emphasize again that the orbifold construction we have considered was originally designed for application to supersymmetric gauge theories. 
The Hamiltonian has similarities to supersymmetric matrix models, which are considered frequently in the context of superstring theory. 
Therefore, this work, together with another work by some of us \cite{matrixModel}, forms a uniform treatment of simulating high energy theories using matrix models. 
To simulate these models on digital quantum hardware, we can use standard Trotter methods (see an example for the Sachdev-Ye-Kitaev (SYK) model \cite{Garcia-Alvarez:2016wem}) or other oracle-based algorithms (for instance, SELECT operators are widely used in quantum chemistry, qubitization, and quantum signal processing, see \cite{mcardle2020quantum} for a comprehensive review). Alternatively, we could apply variational algorithms that are suitable for near-term quantum computers such as the variational quantum eigensolver (for an example of applications of this method to quantum field theories, see \cite{Liu:2020eoa}). Lastly, we could perform analog Hamiltonian simulation in ultracold atomic experiments using, for instance, Rydberg atoms (see, for example, the proposal for simulating the SYK model in \cite{Brown:2019hmk}). Thus, the Hamiltonian formulation of quantum field theories fits naturally into quantum simulation algorithms designed for quantum many-body physics and quantum chemistry. Our work provides a natural step towards solving quantum field theories using quantum devices, ie simulating high energy physics in the lab.
\subsection{Towards gravity/QFT duality}
\hspace{0.51cm}
Since we have discussed quantum simulation of $\text{U}(k)$ gauge theory in this paper, it is natural to address the possibility of studying properties of large-$k$ theory. 
In the context of quantum gravity, this limit is needed to approach the classical gravity regime. 
A natural next step is to formulate protocols for measuring correlation functions involving single, double, or multi-trace operators. 
Taking a proper large-$k$ limit, one can then directly observe the approximate factorization of correlation functions, where the dual gravitational dynamics is semiclassical. Furthermore, one could study scattering problems in AdS/CFT (see, for instance, \cite{Heemskerk:2009pn}) or energy spectra in gauge theories and their relation to quantum gravitational properties in the bulk (see, for instance, problems like~\cite{ElShowk:2011ag}). Such studies will be especially helpful when involving supersymmetry, as in the original work by Maldacena~\cite{Maldacena:1997re}. It would also be interesting to explore the QCD phase diagram in both holographic and non-holographic contexts, where way may study non-trivial phenomena such as confinement-deconfinement phase transitions, Hagedorn behaviors, and holographic Hawking-Page phase transitions~\cite{Witten:1998zw,Sundborg:1999ue,Aharony:2003sx}. 
Direct access to the quantum states in the Hilbert space would also enable us to confirm the recently-proposed microscopic picture of the deconfinement transition~\cite{Hanada:2020uvt,Hanada:2016pwv,Hanada:2018zxn,Hanada:2019czd,Berenstein:2018lrm}. It is promising that the deconfinement transition does not involve very large excitations per color degrees of freedom, as we can check via analytic calculations in weak coupling and lattice Monte Carlo simulations at strong coupling. For example, in the weak-coupling limit of the trivial vacuum of the matrix model, the average excitation level is much less than one (see e.g., Ref.~\cite{Hanada:2019czd} for explicit calculations).  Therefore, a rather small value of the cutoff $\Lambda$ may be enough. The topics mentioned above are important both historically and for the frontier of high energy physics, where quantum simulation could play a significant role. 

\subsection{Adding topological terms}
\hspace{0.51cm}
Finally, we review the possibility of including topological terms. 
Such terms play important roles in quantum field theory, particle physics, and condensed-matter physics. 
In particular, 
so-called $\theta$-terms exist in the standard model of particle physics and
as of yet we have no satisfactory explanation for why the associated $\theta$-angle in QCD is so small in our universe.
This problem is called the strong CP problem. The leading candidate resolution is the axion scenario \cite{Peccei:1977hh} ( see also \cite{Aprile:2020tmw}).
Topological terms in Euclidean space are complex and
therefore the standard approach to simulate QFTs by Markov chain Monte Carlo method suffers 
from the infamous sign problem.
Quantum simulation, however, allows for real-time evolution, which is sign-problem-free \cite{Ortiz:2000gc}. 
Thus, it would be interesting to extend this paper and related results to include topological terms (see a related study in classical computation \cite{Alexandrou:2017hqw} and a recent study about topological terms in quantum simulation \cite{Chakraborty:2020uhf}). 

\section*{Acknowledgement}
\hspace{0.51cm}
We thank Daisuke Kadoh, David B.~Kaplan, Ami Katz, So Matsuura, John Preskill, Fumihiko Sugino, and Mithat Unsal for useful discussions. 
The work of M.~Hanada was supported by the STFC Ernest Rutherford Grant ST/R003599/1. 
He thanks Yukawa Institute for Theoretical Physics for the hospitality during his stay in the summer of 2020. H.G. is supported by the Simons Foundation through the It from Qubit collaboration. 
M.~Honda is partially supported by MEXT Q-LEAP.
JL is supported in part by the Institute for Quantum Information and Matter (IQIM), an NSF Physics Frontiers Center (NSF Grant PHY-1125565) with support from the Gordon and Betty Moore Foundation (GBMF-2644), by the Walter Burke Institute for Theoretical Physics, and by Sandia Quantum Optimization \& Learning \& Simulation, DOE Award \#DE-NA0003525.

\appendix
\section{Orbifold projection from matrix model}\label{sec:orbifold-projection}
\hspace{0.51cm}
In this section, we review the orbifold construction of U($k$) Yang-Mills theory on a 3d spatial lattice, 
with the Lagrangian \eqref{eq:lattice-action}, from a matrix model \cite{Kaplan:2002wv}.  
The same method works for arbitrary dimensions and several other gauge groups. The original motivation for this method was to construct a supersymmetric lattice theory in a systematic manner. However, we demonstrate here that the same process can be used to generate a pure gauge theory without supersymmetry. In principle, one can ignore this derivation and take \eqref{eq:lattice-action} as the starting point. However, we provide it here for clarity.

We begin with the Yang-Mills matrix model with $6$ scalar fields, 
whose Lagrangian is given by 
\begin{eqnarray}
L
=
{\rm Tr}\left(
\frac{1}{2}\sum_{I}(D_tX_I)^2
+
\frac{g_{\rm 1d}^2}{4}\sum_{I,J}[X_I,X_J]^2
\right). 
\end{eqnarray}
The covariant derivative is given by 
\begin{eqnarray}
D_tX_I
=
\partial_tX_I
-
ig_{\rm 1d}[A_t,X_I]. 
\end{eqnarray}
This theory is sometimes called {\it mother theory} in contrast with the {\it daughter theory} obtained by applying the orbifold projection. Following Ref.~\cite{Kaplan:2002wv}, we construct U($k$) Yang-Mills theory with $3$ scalar fields on a $3$-dimensional spatial lattice.
The matrices $X_I$ ($I=1,2,\cdots,6$) are $N\times N$ and Hermitian, where $N=kL^3$ and $L$ will be the length of the spatial lattice. 
We introduce complex matrices $x, y$ and $z$ as
\begin{eqnarray}
x
=
\frac{X_1+iX_2}{\sqrt{2}}, 
\qquad
y
=
\frac{X_3+iX_4}{\sqrt{2}}, 
\qquad
z
=
\frac{X_5+iX_6}{\sqrt{2}}. 
\label{complex-vs-Hermitian}
\end{eqnarray}
Using the notation $\bar{x}=x^\dagger$, $\bar{y}=y^\dagger$ and $\bar{z}=z^\dagger$, the Lagrangian can be written as
\begin{eqnarray}
L
&=&
{\rm Tr}\Biggl(
|D_tx|^2
+
|D_ty|^2
+
|D_tz|^2
-
\frac{g_{\rm 1d}^2}{2}\left|
[x,\bar{x}]
+
[y,\bar{y}]
+
[z,\bar{z}]
\right|^2
\nonumber\\
& &
\qquad
-
2g_{\rm 1d}^2
\left(
|[x,y]|^2
+
|[y,z]|^2
+
|[z,x]|^2
\right)
\Biggl). 
\end{eqnarray}
Here we have used the notation $|M|^2=MM^\dagger$ for any matrix $M$.

\begin{figure}[htbp]
\begin{center}
\scalebox{0.4}{
\includegraphics{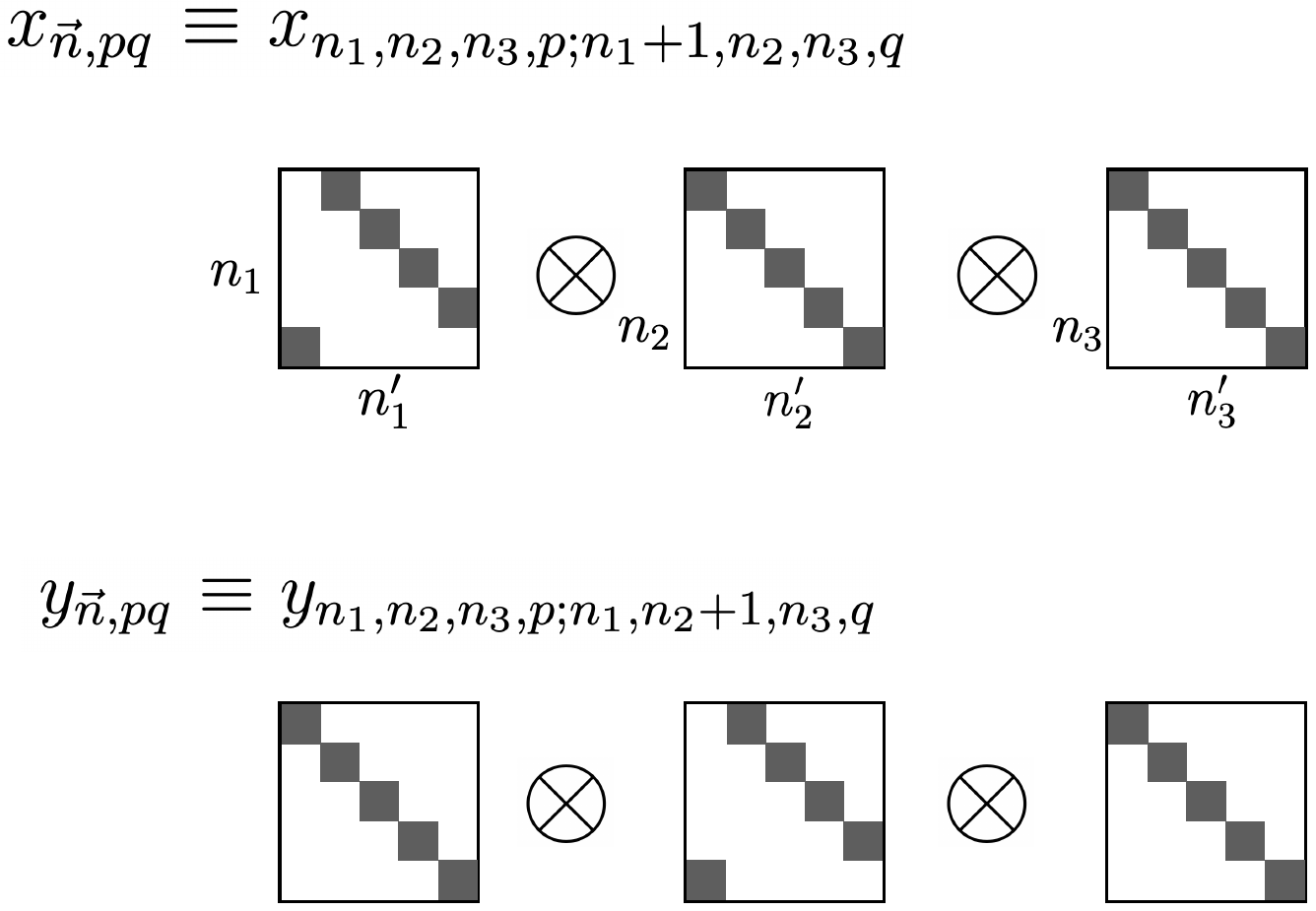}}
\scalebox{0.4}{
\includegraphics{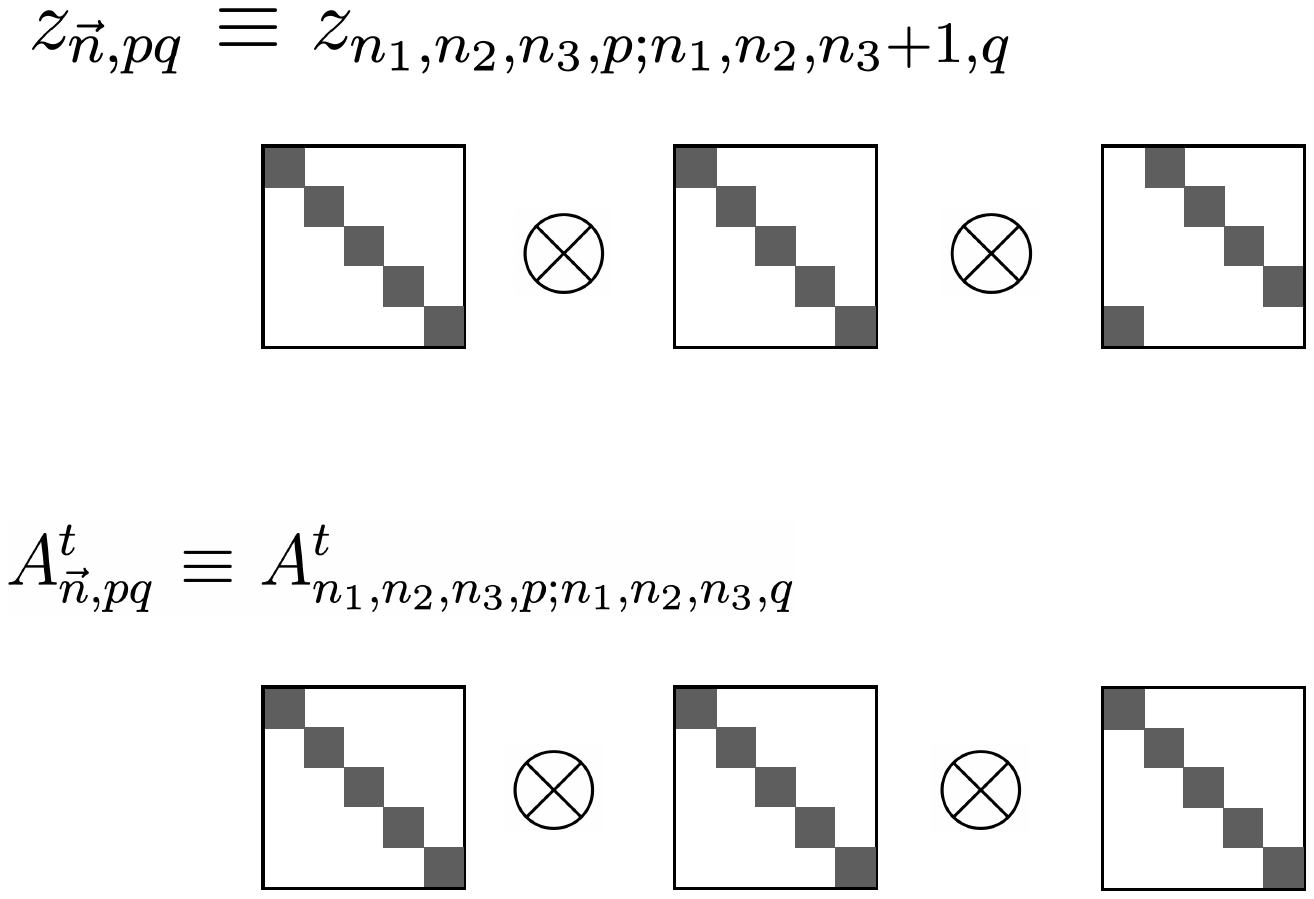}}
\end{center}
\caption{
Schematic picture of the embedding of the orbifold lattice in large matrices. 
Nonzero components \eqref{orbifold_lattice_embedding} are shown in gray. 
}\label{fig:orbifold_lattice}
\end{figure}

We now introduce the so-called `clock' matrices 
\begin{eqnarray}
C_1
&=&
\Omega
\otimes
\textbf{1}_N
\otimes
\textbf{1}_N
\otimes
\textbf{1}_k, 
\nonumber\\
C_2
&=&
\textbf{1}_N
\otimes
\Omega
\otimes
\textbf{1}_N
\otimes
\textbf{1}_k,
\nonumber\\
C_3
&=&
\textbf{1}_N
\otimes
\textbf{1}_N
\otimes
\Omega
\otimes
\textbf{1}_k, 
\end{eqnarray}
where
\begin{eqnarray}
\Omega
=
{\rm diag}
\left(
1,\omega, \omega^2,\cdots,\omega^{L-1}
\right), 
\qquad
\omega
=
e^{-2\pi i/L}. 
\end{eqnarray}
Then, we impose the orbifold projection condition
\begin{align}
C_ixC_i^{-1}
=
\omega^{r_{x,i}}x, 
\qquad
C_iyC_i^{-1}
=
\omega^{r_{y,i}}y, 
\qquad
C_izC_i^{-1}
=
\omega^{r_{z,i}}z,  
\qquad
C_iA_tC_i^{-1}
=
\omega^{r_{A,i}}A_t,  
\end{align}
where
\begin{align}
\vec{r}_x
=
(1,0,0), 
\qquad
\vec{r}_y
=
(0,1,0), 
\qquad
\vec{r}_z
=
(0,0,1),
\qquad
\vec{r}_A
=
(0,0,0). 
\end{align}
To label the matrix entries, we can use $n_{1,2,3},n'_{1,2,3}=1,2,\cdots,L$ and $p,q=1,2,\cdots,k$ instead of $i,j=1,2,\cdots,N=kL^3$, 
respecting the tensor structure of the clock matrices. 
For example, we can use the following convention:
\begin{eqnarray}
& &
x_{ij}
= 
x_{n_1,n_2,n_3,p;n'_1,n'_2,n'_3,q},
\nonumber\\
& &
i=p+(n_1-1)k+(n_2-1)kL+(n_3-1)kL^2,
\nonumber\\
& &
j=q+(n'_1-1)k+(n'_2-1)kL+(n'_3-1)kL^2.
\end{eqnarray}
Then, the only entries surviving after the orbifold projection are
\begin{eqnarray}
x_{\vec{n},pq}
&\equiv&
x_{n_1,n_2,n_3,p;n_1+1,n_2,n_3,q}, 
\nonumber\\
y_{\vec{n},pq}
&\equiv&
y_{n_1,n_2,n_3,p;n_1,n_2+1,n_3,q}, 
\nonumber\\
z_{\vec{n},pq}
&\equiv&
z_{n_1,n_2,n_3,p;n_1,n_2,n_3+1,q}, 
\nonumber\\
A^t_{\vec{n},pq}
&\equiv&
A^t_{n_1,n_2,n_3,p;n_1,n_2,n_3,q}.
\label{orbifold_lattice_embedding}
\end{eqnarray}
See Fig.~\ref{fig:orbifold_lattice}. 
Here periodic boundary conditions are assumed in the notation. 
The crucial step is to interpret $x_{\vec{n}}$, $y_{\vec{n}}$ and $z_{\vec{n}}$ as variables 
on the links connecting $\vec{n}$ and $\vec{n}+\hat{x}$, $\vec{n}+\hat{y}$ and $\vec{n}+\hat{z}$, respectively.  
In this way we identify a `lattice' Lagrangian \eqref{eq:lattice-action}. 

As we have already mentioned, the original motivation of the orbifold construction was to make a lattice with exact supersymmetry.
By taking the mother theory to be a supersymmetric matrix model, the problem of keeping supersymmetry 
reduces to finding a projection condition compatible with supersymmetry.
This is an easier problem than trying to find a supersymmetric lattice without a guiding principle.   
\section{Review of the Kogut-Susskind formulation}\label{sec:Kogut-Susskind}
\hspace{0.51cm}
In this section, we review the Kogut-Susskind formulation \cite{Kogut:1974ag}. 
We consider the (3+1)-d Yang-Mills theory with U($k$) gauge group for simplicity. The generalization to generic gauge groups and spatial dimensions is conceptually straightforward. 

The Kogut-Susskind formulation is a Hamiltonian formulation of lattice gauge theory under the $A_t=0$ gauge. Associated with this gauge choice, the singlet constraint is imposed on the physical states. 

The Hamiltonian is given by the sum of electric and magnetic terms, 
\begin{eqnarray}
\hat{H}
=
\hat{H}_{\rm E}
+
\hat{H}_{\rm B}.  
\end{eqnarray}
The electric term $\hat{H}_{\rm E}$ is
\begin{eqnarray}
\hat{H}_{\rm E}
=
\frac{a^3}{2}
\sum_{\vec{n}}\sum_{\mu=1}^3\sum_{\alpha=1}^{k^2}\left(\hat{E}_{\mu,\vec{n}}^\alpha\right)^2, 
\end{eqnarray}
while the magnetic term $\hat{H}_{\rm B}$ is
\begin{eqnarray}
\hat{H}_{\rm B}
=
-\frac{1}{2ag^2}\sum_{\vec{n}}\sum_{\mu<\nu}
\left(
{\rm Tr}
\left(
\hat{U}_{\mu,\vec{n}}
\hat{U}_{\nu,\vec{n}+\hat{\mu}}
\hat{U}^\dagger_{\mu,\vec{n}+\hat{\nu}}
\hat{U}^\dagger_{\nu,\vec{n}}
\right) + \rm{h.c.}
\right). 
\end{eqnarray}
As usual, the link variable is related to the gauge field as $U_\mu\simeq e^{iagA_\mu^\alpha\tau_\alpha}$, 
where $\tau^\alpha$ ($\alpha=1,2,\cdots,k^2$) are the generators of U($k$), which satisfy 
\begin{eqnarray}
{\rm Tr}\left(
\tau^\alpha\tau^\beta
\right)
=
\delta^{\alpha\beta}, 
\qquad
\sum_{\alpha=1}^{k^2}\tau^\alpha_{pq}\tau^\alpha_{rs}
=
\delta_{ps}\delta_{qr}. 
\end{eqnarray}
The plaquette can be expressed using the field strength tensor $F_{\mu\nu}=\partial_\mu A_\nu-\partial_\nu A_\mu-ig[A_\mu,A_\nu]$
as
$U_{\mu,\vec{n}}U_{\nu,\vec{n}+\hat{\mu}}
U^\dagger_{\mu,\vec{n}+\hat{\nu}}
U^\dagger_{\nu,\vec{n}}=e^{ia^2gF_{\mu\nu}+\cdots}$.
The field strength with spatial indices is the magnetic field: 
$B_1=F_{23}$, $B_2=F_{31}$, $B_3=F_{12}$. 
Hence, in the continuum limit $a\to 0$, $\hat{H}_{\rm E}$ and $\hat{H}_{\rm B}$ as defined above reduce to $\frac{1}{2}\int d^3x{\rm Tr}\hat{\vec{E}}^2$
and $\frac{1}{2}\int d^3x{\rm Tr}\hat{\vec{B}}^2$, respectively. 

The electric field $E^\alpha_\mu$ is identified with $\dot{A}^\alpha_\mu$ (note that we took $A_t=0$), hence 
it should be the conjugate momentum of $A^\alpha_\mu$. 
Therefore, the commutation relation is chosen as 
\begin{eqnarray}
\left[
\hat{E}_{\mu,\vec{n}}^\alpha,
\hat{U}_{\nu,\vec{n}'}
\right]
=
a^{-2}g\delta_{\mu\nu}\delta_{\vec{n}\vec{n}'}\tau_\alpha\hat{U}_{\nu,\vec{n}'}, 
\qquad
\left[
\hat{E}_{\mu,\vec{n}}^\alpha,
\hat{U}^\dagger_{\nu,\vec{n}'}
\right]
=
-a^{-2}g\delta_{\mu\nu}\delta_{\vec{n}\vec{n}'}\hat{U}_{\nu,\vec{n}'}^\dagger\tau_\alpha  
\label{eq:commutation-relation-KS}
\end{eqnarray}
and
\begin{eqnarray}
\left[
\hat{E}_{\mu,\vec{n}}^\alpha,
\hat{E}_{\nu,\vec{n}'}^\beta
\right]
=
-if^{\alpha\beta\gamma}a^{-2}g\delta_{\mu\nu}\delta_{\vec{n}\vec{n}'}\hat{E}^\gamma_{\nu,\vec{n}'}.  
\label{eq:commutation-relation-KS-E}
\end{eqnarray}
Note that, instead of the delta function, $a^{-3}\delta_{\vec{n}\vec{n}'}$ appeared. 
Note also that, once \eqref{eq:commutation-relation-KS} is imposed, 
\eqref{eq:commutation-relation-KS-E} follows because of the Jacobi identity. 
Other commutation relations are 
\begin{eqnarray}
\left[
\hat{U},
\hat{U}
\right]
=
\left[
\hat{U},
\hat{U}^\dagger
\right]
=
\left[
\hat{U}^\dagger,
\hat{U}^\dagger
\right]
=
0.   
\end{eqnarray}

Often, the dimensionless combination $\tilde{E}=a^2g^{-1}E$ is used, 
such that $a^{-2}g$ disappears from the commutation relations, 
\begin{eqnarray}
\left[
\hat{\tilde{E}}_{\mu,\vec{n}}^\alpha,
\hat{U}_{\nu,\vec{n}'}
\right]
=
\delta_{\mu\nu}\delta_{\vec{n}\vec{n}'}\tau_\alpha\hat{U}_{\nu,\vec{n}'}, 
\qquad
\left[
\hat{\tilde{E}}_{\mu,\vec{n}}^\alpha,
\hat{U}^\dagger_{\nu,\vec{n}'}
\right]
=
-\delta_{\mu\nu}\delta_{\vec{n}\vec{n}'}\hat{U}_{\nu,\vec{n}'}^\dagger\tau_\alpha,   
\end{eqnarray}
\begin{eqnarray}
\left[
\hat{\tilde{E}}_{\mu,\vec{n}}^\alpha,
\hat{\tilde{E}}_{\nu,\vec{n}'}^\beta
\right]
=
-if^{\alpha\beta\gamma}\delta_{\mu\nu}\delta_{\vec{n}\vec{n}'}\hat{\tilde{E}}^\gamma_{\nu,\vec{n}'}.  
\end{eqnarray}
The electric part of the Hamiltonian becomes
\begin{eqnarray}
\hat{H}_{\rm E}
=
\frac{g^2}{2a}
\sum_{\vec{n}}\sum_{\mu=1}^3\sum_{\alpha=1}^{k^2}\left(\hat{\tilde{E}}_{\mu,\vec{n}}^\alpha\right)^2.
\end{eqnarray}

In order to simplify the notation in the large-$k$ limit, we can change the normalization of $\tau_\alpha$ as
$\tau'\equiv k^{-1}\tau$, 
\begin{eqnarray}
{\rm Tr}\left(
\tau^{\prime\alpha}\tau^{\prime\beta}
\right)
=
\frac{\delta^{\alpha\beta}}{k^2}, 
\qquad
\sum_{\alpha=1}^{k^2}\tau^{\prime\alpha}_{pq}\tau^{\prime\alpha}_{rs}
=
\frac{\delta_{ps}\delta_{qr}}{k^2}. 
\end{eqnarray}
Then by rescaling $f_{\alpha\beta\gamma}$ and $\tilde{E}^\alpha$ as 
$f'_{\alpha\beta\gamma}=k^{-1}f_{\alpha\beta\gamma}$ and 
$\tilde{E}^{\prime\alpha}=k^{-1}\tilde{E}^\alpha$ 
we can write the commutation relation as 
\begin{eqnarray}
\left[
\hat{\tilde{E}}_{\mu,\vec{n}}^{\prime\alpha},
\hat{U}_{\nu,\vec{n}'}
\right]
=
\delta_{\mu\nu}\delta_{\vec{n}\vec{n}'}\tau'_\alpha\hat{U}_{\nu,\vec{n}'}, 
\qquad
\left[
\hat{\tilde{E}}_{\mu,\vec{n}}^{\prime\alpha},
\hat{U}^\dagger_{\nu,\vec{n}'}
\right]
=
-\delta_{\mu\nu}\delta_{\vec{n}\vec{n}'}\hat{U}_{\nu,\vec{n}'}^\dagger\tau'_\alpha,   
\end{eqnarray}
\begin{eqnarray}
\left[
\hat{\tilde{E}}_{\mu,\vec{n}}^{\prime\alpha},
\hat{\tilde{E}}_{\nu,\vec{n}'}^{\prime\beta}
\right]
=
-if^{\prime\alpha\beta\gamma}\delta_{\mu\nu}\delta_{\vec{n}\vec{n}'}\hat{\tilde{E}}^{\prime\gamma}_{\nu,\vec{n}'}.  
\end{eqnarray}
The electric part of the Hamiltonian becomes
\begin{eqnarray}
\hat{H}_{\rm E}
=
\frac{\lambda k}{2a}
\sum_{\vec{n}}\sum_{\mu=1}^3\sum_{\alpha=1}^{k^2}\left(\hat{\tilde{E}}_{\mu,\vec{n}}^{\prime\alpha}\right)^2. 
\end{eqnarray}
Here $\lambda=g^2k$ is known as the 't Hooft coupling. 
The magnetic part is 
\begin{eqnarray}
\hat{H}_{\rm B}
=
-\frac{k}{2a\lambda}\sum_{\vec{n}}\sum_{\mu<\nu}
\left(
{\rm Tr}
\left(
\hat{U}_{\mu,\vec{n}}
\hat{U}_{\nu,\vec{n}+\hat{\mu}}
\hat{U}^\dagger_{\mu,\vec{n}+\hat{\nu}}
\hat{U}^\dagger_{\nu,\vec{n}}
\right) +\rm{h.c.}
\right). 
\end{eqnarray}
With this convention, it is clear that in the `strong coupling limit' $\lambda\to\infty$, the magnetic term is omitted.

The operator $\hat{U}_{\mu,\vec{n}}$ is interpreted as the coordinate of the group manifold U($k$) for the link variable on the site $\vec{n}$ in the $\mu$-direction. Ignoring the gauge-singlet constraint, the Hilbert space is formally written as 
\begin{eqnarray}
{\cal H}
=
\otimes_{\mu,\vec{n}}{\cal H}_{\mu,\vec{n}}
\sim
\otimes_{\mu,\vec{n}}
\left(
\oplus_{g\in{\rm U}(k)}
|g\rangle_{\mu,\vec{n}}
\right), 
\end{eqnarray}
 where
 \begin{eqnarray}
\hat{U}_{\mu,\vec{n}}
|g\rangle_{\mu,\vec{n}}
=
g|g\rangle_{\mu,\vec{n}}. 
\end{eqnarray}
More precisely, we will consider only the Hilbert space of square-integrable wave functions on U($k$):
\begin{eqnarray}
|f\rangle
=
\int_G dgf(g)|g\rangle, 
\qquad
\int_G dg |f(g)|^2<\infty,
\end{eqnarray}
where we use the Haar measure for the integration. 
In other words, ${\cal H}_{\mu,\vec{n}}=L^2(G)$, where $L^2(G)$ is the set of square-integrable functions from $G$ to ${\mathbb C}$. Physically, 
this means that we consider only the normalizable states.  
\subsection{Realization on a quantum computer}
\hspace{0.51cm}
How should we regularize this Hilbert space systematically? 
It would be nice if the group manifold could be discretized by a discrete subgroup, but this does not seem to work
except for U(1) theory, where ${\mathbb Z}_N$ gives an efficient discretization.\footnote{
See Ref.~\cite{Alexandru:2019nsa} for attempts to use a large discrete subgroup of SU(3). } 

A physically elegant, but practically very hard, approach is to truncate in the loop basis. 
We start with the `strong coupling limit,' where $\hat{H}_{\rm B}$ is dropped. 
The ground state in this limit is given by $\hat{E}_{\mu,\vec{n}}^\alpha|0\rangle=0$. 
By acting on this state with Wilson loop operators $\hat{W}_{C}$, obtained by multiplying 
the link variables along a closed contour $C$ and taking their trace, for various contours, 
an over-complete basis of the gauge-invariant Hilbert space is obtained.
The magnetic term is regarded as the smallest Wilson loop, i.e., the plaquette.   
When the loops do not intersect with each other or with themselves, 
the electric term is proportional to the sum of the lengths of the loops. 
When the loops intersect, the electric term joins or splits them. 
By identifying the Wilson loop with the string, this gives an alternative picture to the unitary link variables. 
A natural cutoff can be introduced by restricting the total length of a string,
however, there is no known way to write down an orthonormal basis for the physical states systematically. 

For quantum computation, probably the most natural option is to use the Peter-Weyl theorem, 
which gives the ``Fourier expansion" on group manifolds:
\\
\vspace{-2mm}
\\
{\bf Peter-Weyl theorem.} For a compact group $G$, an orthonormal basis of $L^2(G)$ is given by the matrix coefficients of the unitary, finite-dimensional irreducible representations (irreps) of $G$, $\rho^{(R)}_{ij}$, where 
$R$ runs through all irreps, and $i,j=1,2,\cdots,{\rm dim}R$. 
\\
\vspace{-2mm}
\\
In particular, $f\in L^2(G)$ can be written as $f(g)=\sum_R \sum_{i,j=1}^{{\rm dim}R}c^{(R)}_{ij}\rho^{(R)}_{ij}(g)$ for $g \in G$. 
There are two canonical orthonormal bases on $L^2(G)$; in bra-ket notation, $\{|g\rangle\}$ is the `coordinate' basis of group elements $g$, and $\{|R,ij\rangle\}$ is the `momentum' basis provided by the Peter-Weyl theorem. We refer to $\langle g|R,ij\rangle=\rho^{(R)}_{ij}(g)$ as a `Fourier mode' on the group $G$. Accordingly, the Hilbert space can be expressed as 
\begin{eqnarray}
{\cal H}
=
\otimes_{\mu,\vec{n}}{\cal H}_{\mu,\vec{n}}
=
\otimes_{\mu,\vec{n}}
\left(
\oplus_{R}
\oplus_{i,j=1}^{{\rm dim}R}
|R,ij\rangle_{\mu,\vec{n}}
\right). 
\end{eqnarray}
Note that $i$ and $j$ in $|R,ij\rangle_{\mu,\vec{n}}$ are transformed by gauge transformations at sites $\vec{n}$
and $\vec{n}+\hat{\mu}$. A natural cutoff is introduced by restricting to a subset of the representations which comprise the Hilbert space \cite{Zohar:2014qma}. 

When acting on $|R,ij\rangle$, the electric term
$\hat{H}_{\rm E}$ is proportional to the quadratic Casimir operator,
\begin{equation}
\hat{H}_{\rm E}|R,ij\rangle \propto \chi^2(R)|R,ij\rangle.
\end{equation}
The action of the magnetic term $\hat{H}_{\rm B}$ is more complicated:
\begin{eqnarray}
\hat{U}_{pq}|R,ij\rangle
&=&
\int_G dg
\sum_{R',i',j'}
|R',i'j'\rangle\langle R',i'j'|
\hat{U}_{pq}|g\rangle\langle g|R,ij\rangle
\nonumber\\
&=&
\sum_{R',i',j'}
|R',i'j'\rangle
\int_G dg\left(\rho^{(k)}_{pq}(g)\rho^{(R)}_{ij}(g)\left(\rho^{(R')}_{i'j'}(g)\right)^\ast\right)
\nonumber\\
&=&
\sum_{R',i',j'}
C_{R'i'j';Rij;k,pq}
|R',i'j'\rangle, 
\end{eqnarray}
so that
\begin{eqnarray}
\langle R',i'j'|
\hat{U}_{pq}|R,ij\rangle
=
C_{R'i'j';Rij;k,pq}, 
\end{eqnarray}
where $C_{R'i'j';Rij;k,pq}\equiv\langle R',i'j'|\cdot\left(|R,ij\rangle\otimes|k,pq\rangle\right)$ is the generalized version of the Clebsch-Gordan coefficient
and `$k$' means that $\rho^{(k)}_{pq}(g)$ is a $k\times k$ matrix, i.e.~the fundamental representation of U($k$).

As a natural way to regularize the Hilbert space, we can introduce a cutoff for the dimension of the representations. This can be interpreted as the momentum cutoff on the group manifold. 
\subsection*{Preparation of ground state}
\hspace{0.51cm}
Let us define $\hat{H}(s)$ as 
\begin{eqnarray}
\hat{H}(s)
=
\hat{H}_{\rm E}
+
s\hat{H}_{\rm B},   
\end{eqnarray}
where $0\le s\le 1$.
At $s=0$, $\hat{H}(s=0)=\hat{H}_{\rm E}$ is the `strong coupling limit'; the ground state is given by a vanishing electric field, i.e.~the trivial representation, on every link. 
Starting from this trivial limit, we can apply the adiabatic state preparation method to prepare the ground state of $\hat{H}_{\rm E} + \hat{H}_{\rm B}$. 
As with the orbifold construction, it is important to look at the gap as a function of $s$ in the extended Hilbert space containing the gauge-non-singlet modes to fully understand the complexity of the adiabatic state preparation procedure.

\section{Regularization in the coordinate basis}\label{sec:coordinate_basis}
\hspace{0.51cm}
In this Appendix, we introduce a regularization in the coordinate basis~\cite{Jordan:2011ne,Jordan:2011ci,Klco:2018zqz}. 
Let $\{\ket{x}\}$ be the coordinate basis for a particle in flat space, which satisfies
\begin{eqnarray}
\hat{x}\ket{x}
=
x\ket{x}. 
\end{eqnarray}
The simplest way to regularize it is to introduce the cutoff to the value of $x$ as 
\begin{eqnarray}
-R\le x\le R, 
\end{eqnarray}
and introduce $\Lambda$ lattice points, 
\begin{eqnarray}
x_n
=
-R+n\delta_x,
\qquad
\delta_x=\frac{2R}{\Lambda-1}, 
\qquad
n=0,1,\cdots,\Lambda-1  
\end{eqnarray}
The regularization parameters $\Lambda$, $\delta_x$ and $R$ should be sent to infinity, zero and infinity, respectively. Roughly speaking, $\delta_x$ and $R$ correspond to $\mu$ and $\omega$ in the regularization scheme introduced in Sec.~\ref{sec:Fock-space-truncation}. By using $\ket{n}$ to denote $\ket{x_n}$, we can write
\begin{eqnarray}
\hat{x}
=
\sum_{n=0}^{\Lambda-1}
x_{n}
\ket{n}\bra{n}. 
\end{eqnarray}
By using the binary decomposition as in Sec.~\ref{sec:Fock-space-truncation}, we can rewrite it to a sum of the Pauli strings. 

The momentum operator $\hat{p}$ appears in the Hamiltonian only in the form of $\hat{p}^2$; 
a convenient way of regularizing it is 
\begin{eqnarray}
\hat{p}^2
=
\frac{1}{\delta_X^2}
\sum_{n=0}^{\Lambda-1}
\left\{
2\ket{n}\bra{n}
-
\ket{n+1}\bra{n}
-
\ket{n}\bra{n+1}
\right\}. 
\label{p^2-coordinate-basis}
\end{eqnarray}
This form can be understood as follows. Above, we introduced the cutoff for the value of $x$, but we could use the periodic boundary condition $\ket{\Lambda}=\ket{0}$ as well, assuming that the states close to the cutoff do not give non-negligible contributions. In this case, the `shift operator' $\hat{S}\equiv\sum_n\ket{n+1}\bra{n}$ is identified with $e^{i\delta_X\hat{p}}$, and hence,   
$\hat{p}=\frac{\hat{S}^{1/2}-\hat{S}^{-1/2}}{i\delta_X}$, up to the corrections of order $\delta_X$. 
From this, $\hat{p}^2=\frac{2\hat{I}-\hat{S}-\hat{S}^{-1}}{\delta_X^2}$ follows. 
This is the same as \eqref{p^2-coordinate-basis} up to the boundary condition. 
Again, it is straightforward to write the right-hand side of \eqref{p^2-coordinate-basis} as a sum of Pauli strings, upon which efficient simulation algorithms may be applied.

\bibliographystyle{utphys}
\bibliography{orbifold}
\end{document}